\documentclass[usenatbib]{mnras}

\usepackage{placeins}
\usepackage{euscript}
\usepackage{booktabs}

\usepackage[T1]{fontenc}

\DeclareRobustCommand{\VAN}[3]{#2}
\let\VANthebibliography\thebibliography
\def\thebibliography{\DeclareRobustCommand{\VAN}[3]{##3}\VANthebibliography}

\usepackage{newtxtext,newtxmath}

\usepackage{ulem}
\usepackage{subcaption}
\usepackage{tablefootnote}

\usepackage[dvipsnames]{xcolor}
\usepackage{comment}

\usepackage{graphicx}	
\usepackage{amsmath}	
\usepackage{hyperref}
\usepackage{ulem,xcolor}

\usepackage{multirow}

\usepackage{enumitem}
\usepackage{soul,color}

\usepackage{threeparttable}
\defcitealias{2025MNRAS.543.2180R}{R25}
\usepackage{caption}
\title[1.4 GHz diagnostics for SN Ia hosts]{Applications of 1.4 GHz diagnostics to Type Ia Supernova host galaxies}

\author[S. Ramaiya et al.]{
S. Ramaiya,$^{1}$\thanks{E-mail: shruti.ramaiya@physics.ox.ac.uk}
M. J. Jarvis,$^{1,2}$
M. Vincenzi,$^{1}$
M. Sullivan,$^{3}$
I. H Whittam$^{1,2}$
\\
$^{1}$Astrophysics, Department of Physics, University of Oxford, Keble Road, Oxford, OX1 3RH, UK\\
$^{2}$Department of Physics and Astronomy, University of the Western Cape, Robert Sobukwe Road, 7535 Bellville, Cape Town, South Africa\\
$^{3}$School of Physics and Astronomy, University of Southampton, Southampton, SO17 1BJ, UK
}

\date{Accepted XXX. Received YYY; in original form ZZZ}

\pubyear{\the\year{}}

\begin{document}
\label{firstpage}
\pagerange{\pageref{firstpage}--\pageref{lastpage}}
\maketitle

\begin{abstract}

Type Ia supernova (SN~Ia) standardisation parameters exhibit evidence for systematic variation across the host galaxy star-formation rate--stellar mass (SFR$-M_\star$) plane, motivating the incorporation of galaxy SFR information in  cosmological inference. SFRs are commonly estimated via spectral energy distribution (SED) fitting with far-infrared (FIR) measurements to account for dust-obscured star formation. Such FIR coverage will, however, be limited for upcoming time-domain surveys such as the Rubin Observatory Legacy Survey of Space and Time (LSST), necessitating the use of alternative SFR tracers. Here, we reconstruct the SFR--$M_\star$ plane using 1.4 GHz diagnostics, to test the consistency of host classifications against FIR-constrained SED-based estimates. Within this plane, SN Ia host galaxies are divided into three regions: Region~1 (low-mass), Region~2 (high-mass star-forming) and Region~3 (high-mass passive). We find that ${\sim}84$ per cent of SN hosts retain identical region assignments when using radio versus FIR-constrained SED-derived SFRs. Measuring SN~Ia nuisance parameters ($\alpha,\beta, M$) within each subregion, we find consistent values between the two SFR--$M_\star$ plane reconstructions, indicating limited sensitivity to SFR estimator choice, with the largest deviations in Region~3 at ${\sim}1.1\sigma$. Across the three 1.4 GHz SFR--$M_\star$ subregions, we confirm the region-dependent variation in SN~Ia standardisation parameters--particularly $\beta$--reported in our earlier SED-based analysis. With near-complete radio coverage of the LSST footprint anticipated from current and forthcoming radio continuum surveys (e.g., Square Kilometre Array), radio SFR calibrations will become an increasingly useful and scalable approach to host galaxy classification, supporting the construction of robust SN Ia subsamples for precision cosmology.

\end{abstract}


\section{Introduction}

Type Ia supernovae (SNe Ia) are the most direct probe of the cosmic expansion history and provide key constraints on its underlying dark energy component. In the advent of wide-field programs such as the Vera Rubin Observatory Legacy Survey of Space and Time (LSST; \citealp{2019ApJ...873..111I}) and the Nancy Grace Roman Space Telescope (\citealp{2018ApJ...867...23H,2021arXiv211103081R}), the coming decade ushers in a golden era of transient discovery, with observations of over a million SNe expected. While this discovery rate is exciting, realising the full scientific potential of these datasets will require a coordinated effort to gather detailed environmental information for all confirmed SNe Ia. Host characterisation is crucial as there is extensive evidence that SN Ia luminosities depend on their galaxy properties, with correlations observed between host stellar mass (\citealp{2010ApJ...715..743K,2010MNRAS.406..782S,2010ApJ...722..566L}), metallicity \citep{2011ApJ...743..172D,2013ApJ...770..108C} and star formation rate (SFR; \citealp{2006ApJ...648..868S,2010MNRAS.406..782S}) among others. A complete picture of the physical mechanisms driving these correlations, however, remains lacking, highlighting the need for continued, targeted investigation.

Informed by galaxy evolution studies, the SFR--$M_\star$ plane (e.g., \citealp{2007ApJ...660L..43N,2007A&A...468...33E,2007ApJ...670..156D,2012ApJ...754L..29W,2015MNRAS.453.2540J,2016MNRAS.461..458D}) provides a key parameter space to study SN Ia--host correlations. The bulk of star-forming galaxies lie along a narrow sequence in this plane, known as the galaxy star-forming main sequence (SF-MS). This relation is observed across a wide range of redshifts $(0 \lesssim z \lesssim 4)$ and is typically characterised by an approximately unity slope, with an intrinsic scatter of ${\sim}0.2-0.3$ dex. Beyond this dominant self-regulated regime, the SFR--$M_\star$ plane encompasses a broader population of galaxies, ranging from elevated `starbursts' to quiescent systems with suppressed activity. Such diversity arises from the cumulative effects of physical processes that regulate star formation, such as gas supply \citep{2006MNRAS.367.1394K,2008A&ARv..15..189S}, mergers (e.g., \citealp{2009MNRAS.394.1956C,2009ApJ...697.1369B,2011ApJ...742..103L}), feedback \citep{2004MNRAS.353..713K,2008MNRAS.387.1431D,2008MNRAS.389.1137S,2012ARA&A..50..455F} and quenching \citep{2010ApJ...721..193P,2014MNRAS.440..889S,2016ApJ...825..113D,2019MNRAS.483.5444D}. A galaxy's position in this space therefore encodes information about its evolutionary state and provides a natural framework for defining physically distinct sub-populations.

Accurate measurements of the constituent observables are essential for studies of the SFR--$M_\star$ plane. While galaxy stellar masses can be estimated with relatively well-understood systematics, measuring SFRs remains considerably more challenging (see \citealp{1998ARA&A..36..189K,2012ARA&A..50..531K} for reviews). Common star formation diagnostics, such as UV continuum and nebular emission lines (e.g., H$\alpha$), are subject to systematic limitations arising from dust attenuation and aperture-related effects \citep{2001ApJ...558...72S,2003ApJ...599..971H, 2011MNRAS.415.1647G}. Infrared (IR)-based SFR tracers, by contrast, are largely insensitive to dust obscuration, as they trace dust-reprocessed stellar emission. However, in regions of largely unobscured star-formation (e.g., low-dust or metal-poor systems), IR emission can provide an incomplete measure of the total SFR (e.g., \citealp{2007ApJ...666..870C,2010ApJ...714.1256C,2009ApJ...703.1672K}). To mitigate the limitations from single-band diagnostics, composite SFR indicators that leverage information across multiple wavelengths---such as hybrid UV--IR measurements (e.g., \citealp{2000ApJ...533..236G,2003ApJ...586..794B,2003A&A...410...83H,2006ApJS..164...38I,2008MNRAS.386.1157C,2011ApJ...741..124H}) or full spectral energy distribution (SED) fitting (e.g., \citealp{2008MNRAS.388.1595D,2009A&A...507.1793N,2023ApJ...944..141P})---are frequently adopted in the literature. By jointly capturing the unobscured emission from newly formed stars and the dust-reprocessed component, this approach removes the need for explicit dust–obscuration corrections.

SED fitting analyses of SN Ia host galaxies have historically relied almost exclusively on deep optical broadband imaging from SN search and follow-up programmes. In the subset of studies that have placed SN Ia host galaxies on the SFR–$M_\star$ plane (e.g., \citealp{2006ApJ...648..868S,2010ApJ...722..566L,2020MNRAS.494.4426S}), the SED constraints are drawn exclusively from \textit{ugriz} photometry. In a later work, \cite{2010MNRAS.406..782S} incorporate additional near-infrared (NIR) \textit{JHK} photometry into their SFR–$M_\star$ plane analysis of SN Ia host galaxies where available. It is well known, however, that physical parameters inferred from SED fitting of optical/NIR photometry alone are affected by strong degeneracies (see \citealp{2013ARA&A..51..393C} for a review). The ``age--metallicity--dust'' degeneracy (e.g., \citealp{2001ApJ...559..620P}), in particular, makes it difficult to reliably measure dust attenuation, and hence SFRs, unless far-infrared (FIR) data are available. This uncertainty can propagate into galaxy classifications, complicating the separation of passive galaxies from dust-obscured, star-forming systems. When studying SN Ia subsamples in this context, any misestimation of host galaxy properties can, in turn, lead to spurious SN Ia-host correlations or mask genuinely physical trends.

In \citet[hereafter Paper I]{2025MNRAS.543.2180R}, we model the SEDs of SN Ia host galaxies in the Dark Energy Survey (DES; \citealp{2005astro.ph.10346T}) using photometry extending to the FIR. At the time of writing, this offers the widest wavelength coverage used in studies of SN~Ia hosts. We show that the mid-/FIR data is crucial for mitigating degeneracies in SED-derived SFRs and can yield significantly different estimates compared to optical-only analyses. Using these improved host galaxy property estimates, SNe~Ia  are classified into three distinct groups according to the distribution of their host galaxies on the SFR--$M_\star$ plane. Each region comprises galaxies at different evolutionary stages: Region 1 -- low-mass ($<~10^{10}$ $\text{M}_\odot$) hosts, Region 2 -- high-mass, star-forming hosts and Region 3 -- high-mass, passive hosts. We find the colour-luminosity slope, `$\beta$' (a SN~Ia standardisation parameter), is steepest for SNe~Ia in Region 1 hosts ($\beta = 3.51 \pm 0.16$) and shallowest in Region 3 ($\beta = 2.12 \pm 0.16$), differing at the ${\sim}6\sigma$ level. After correcting each subsample by its respective $\beta$, Region 3 SNe~Ia (high-mass, passive hosts) appear 0.07 -- 0.12 mag $(>3\sigma)$ brighter, post-standardisation. These findings suggest that future cosmological analyses should apply standardisation relations to SNe~Ia according to the region of the SFR--$M_\star$ plane their host galaxies occupy.

In this paper, we consider the feasibility of recreating the analysis in Paper~I on the scale of LSST and other future time-domain surveys. In such \textit{optical-only} survey regimes, supplementary longer-wavelength data from other facilities will be needed for measurements of SED-based SFRs. The long-wavelength data in Paper~I comes from the \textit{Herschel} and \textit{Spitzer} space telescope missions. However, within the 18,000 deg$^2$ LSST observing footprint, deep \textit{Herschel} and \textit{Spitzer} coverage is limited to a few ${\sim}100$ deg$^2$. For regions of the LSST footprint that lack mid-/FIR coverage, an alternate prescription is required to robustly measure SFRs and break degeneracies among host galaxy parameters.

One such alternative is the radio continuum, which offers a dust-unbiased tracer of star-formation. Such calibrations are bootstrapped from a well-known empirical correlation between the FIR and 1.4 GHz radio emission (\citealp{1985A&A...147L...6D,1985ApJ...298L...7H,1992ARA&A..30..575C,2001ApJ...554..803Y,2003ApJ...586..794B,2010MNRAS.409...92J,2011ApJ...737...67M, 2017MNRAS.466.2312D, 2024MNRAS.531..708C}). The FIR--radio correlation (FIRC) holds over many orders of magnitude and exhibits a remarkably tight scatter ($\sim$0.3 dex), leading to the conclusion that emission at both wavelengths share a common origin (massive stars) and trace recent star-formation in the local and distant universe. The FIR emission comes mostly from young, massive $(>5\hspace{0.2em}\text{M}_\odot)$ type O/B stars that heat the dust in their surrounding birth clouds, which then re-radiate in the infrared. As for the radio emission at 1.4 GHz, this is mainly (non-thermal) synchrotron emission generated by  cosmic ray electrons that have been accelerated by SN explosions of massive stars $(>8\hspace{0.2em}\text{M}_\odot)$ \citep{1992ARA&A..30..575C,2009ApJ...706..482M}. The integrated radio luminosity also has a thermal component due to free-free emission from HII regions but, at 1.4 GHz, this contribution is largely negligible $({\sim}10\%;$ \citealp{1992ARA&A..30..575C}$)$.

Radio-based star-formation calibrations will become increasingly important in the LSST era, as deep radio continuum surveys—both current and forthcoming—provide significant overlap with optical imaging. The Evolutionary Map of the Universe \citep[EMU; ][]{EMU,EMU2025} will provide near-complete overlap with the LSST footprint, surveying the entire Southern hemisphere at 1.3~GHz and extending to $\delta \simeq +30^\circ$, with completion expected by 2028. In this wide-area tier, EMU together with new MeerKAT continuum programmes \citep[e.g.][]{MeerKLASScont} will reach sufficient sensitivity to detect star-forming galaxies out to $z\sim 0.5$. In the deep drilling fields, the MeerKAT International GHz Tiered Extragalactic Exploration (MIGHTEE) survey \citep{2016mks..confE...6J, 2022MNRAS.509.2150H,2025MNRAS.536.2187H} has already demonstrated the power of deep radio data to probe the properties and evolution of star-forming galaxies to $z\sim 5$ \citep{2025ApJ...989...44T,2025MNRAS.543..507W,Thykkathu2026, 2026arXiv260205808V}. Looking further ahead, the Square Kilometre Array Observatory (SKAO) will extend this ``wedding-cake'' survey strategy to even greater depth and angular resolution \citep[e.g.][]{2015aska.confE..67P,2015aska.confE..68J}, producing radio continuum data that will detect typical SFGs to $z\sim 1$ over thousands of square degrees. Ahead of these next-generation surveys, it is vital to use existing deep radio data sets and calibrations to test the reliability of radio luminosity diagnostics as an independent SFR estimator (e.g., \citealp{2023A&A...675A.126A}) for SN Ia host galaxies.

A plan of the paper is as follows: In Section~\ref{sec: data} we introduce the SN~Ia sample and the radio data available for their host galaxies. The methodology used to derive 1.4 GHz luminosities and the selection cuts applied to obtain the final sample are also outlined. Section~\ref{sec:umbrella} explores the application of radio-based star-formation diagnostics in the context of SN Ia host galaxies. We present SFRs derived from SED fitting as a function of 1.4 GHz luminosity for the SN~Ia host galaxies and show these data relative to radio--SFR calibrations from the literature. Adopting the calibration of \cite{2024MNRAS.531..708C}, we construct the SFR--$M_\star$ plane, as done in Paper~I, but using 1.4 GHz SFRs in place of SED-derived estimates. This allows us to assess how consistently galaxies are assigned to the three regions identified in Paper~I and to study SN~Ia subsamples defined by each region. Predictions surrounding the future use of radio survey data are given in Section \ref{sec:predictions}. We summarise and conclude in Section \ref{sec:conclusions}.

Throughout the paper, where relevant, we assume a flat $\Lambda$CDM cosmological model 
with $\Omega \rm_M = 0.315$, $\Omega_\Lambda = 0.685$ and $H_0 = 70$ km s$^{-1}$ Mpc$^{-1}$ (following Planck Collaboration VI \citeyear{2020A&A...641A...6P}). The radio spectral index is taken to have a value, $\alpha_{\rm rad} = 0.7$\footnote{By convention, we define the spectral index such that the radio flux density scales with frequency as $S_\nu \propto \nu^{-\alpha_{\rm rad}}$.}, assuming that the integrated radio emission is synchrotron-dominated at 1.4 GHz.

\section{Data and methodology overview} \label{sec: data}

A description of the SN Ia data sets and procedures is given in Paper~I of this series. In this section, we briefly summarize the SN~Ia sample for completeness and introduce the associated host galaxy radio data and analysis methods.

\subsection{Supernova data}

Our SN~Ia data come from the Dark Energy Survey (DES). DES is an optical imaging survey that ran for a period of six years and covered {$\sim$}5100 deg$^2$ of the Southern hemisphere using the Dark Energy Camera \citep[DECam;][]{2015AJ....150..150F} on the Victor M. Blanco 4-m telescope (Cerro Tololo, Chile). DES was designed to constrain the properties of dark energy using four complementary probes: galaxy clusters, weak gravitational lensing, large scale structure and SNe~Ia. For time-domain science, the DES Supernova (DES-SN) programme monitored ten 2.7-deg$^2$ fields in \textit{griz} with a typical ${\sim}7$-day observer frame cadence. Eight (C1, C2, E1, E2, S1, S2, X1, X2) were `shallow' fields, surveyed to a depth of ${\sim}23.5$ mag per epoch, and two (C3, X3) were `deep fields', with a depth of ${\sim}24.5$ mag. This paper uses data from `DES-SN5YR' \citep{2024ApJ...975....5S}\footnote{\label{desdr}\url{https://github.com/des-science/DES-SN5YR}} which comprises photometrically classified SNe~Ia from the full five years of DES, for which spectroscopic redshifts are available for the host galaxies. Selection cuts are applied as described in Paper~I to yield our primary `gold' sample consisting of 501 SN~Ia host galaxies. 

\subsubsection{Distance estimation}

SN Ia distance moduli, $\mu_{\text{obs}}$ are estimated using (e.g., \citealp{1998A&A...331..815T,2006A&A...447...31A}),

\begin{equation} \label{eq:distmod}
    \mu_{\text{obs}} = m_x +\alpha x_1 -\beta c -M -\Delta \mu_\text{bias} ,
\end{equation}
where $m_x$, $x_1$ and $c$ are the SN~Ia light-curve parameters as described in the SALT3 model framework \citep{2007A&A...466...11G,2021ApJ...923..265K}, representing the light-curve amplitude, stretch and colour, respectively. The light curve fit parameters used in this analysis come from the DES-SN5YR data release\footref{desdr} \citep{2024ApJ...975....5S}. The nuisance parameters $\alpha$ and $\beta$ parametrise the stretch-luminosity and colour-luminosity relations, respectively, and $M$ is the absolute magnitude of a SN Ia with $x_1 = 0$ and $c =0$. Biases arising from various selection effects and choices in analysis are accounted for using the $\Delta \mu_\text{bias}$ term. Standard cosmological analyses include an additional term that accounts for any residual dependencies between standardised (i.e. colour- and stretch-corrected) SN~Ia luminosities and their host galaxy properties. This is typically denoted by $\gamma G_\text{host}$ and defined as a step function of the form,

\begin{equation}
    \gamma G_{\mathrm{host}} = \begin{cases}
        +\gamma/2 & P > P_{\mathrm{step}}, \\
    -\gamma/2 & \mathrm{otherwise}, 
    \end{cases}
\end{equation}

\noindent where $\gamma$ is the residual `step' size, $P$ is a chosen property of the SN host galaxy and $P_{\mathrm{step}}$ is the threshold value defining the step. It is usual in cosmological analyses of SNe~Ia to take the stellar mass as the host galaxy property, with the step measured on either side of a 10$^{10}\rm M_\odot$ threshold (i.e., the ``mass step''; \citealp{2010ApJ...715..743K,2010MNRAS.406..782S,2010ApJ...722..566L}). In the interest of studying correlations between SNe Ia and their host galaxies, we do not use this term in our analysis. Here, SNe~Ia in low- and high-mass galaxies are fit \textit{separately} and therefore this mass step term is simply absorbed by $M$. 

We further note that the robustness of our results against selection effects was explored in our earlier work using both the full DES-SN sample (up to $z\sim1.2$) and a redshift-restricted sample ($z<0.6$),  where such effects are expected to be significantly reduced. This test is not repeated here and we refer the reader to Paper~I (their Section 2; results in their Table~4)

\subsection{Galaxy data}

MeerKAT is a 64-dish radio interferometer located in South Africa, with each receptor comprising a 13.5-m diameter main reflector and a 3.8-m diameter sub-reflector. The telescope is equipped with receivers operating across three frequency bands: UHF--band ($580 < \nu <1015$ MHz), $L$--band ($900 < \nu < 1670$ MHz) and S--band ($1750 < \nu < 3500$ MHz). Our radio continuum data comes from the MeerKAT International GHz Tiered Extragalactic Exploration (MIGHTEE; \citealp{2016mks..confE...6J}) survey, which is one of the Large Survey Projects currently being conducted with MeerKAT. It is mapping ${\sim}20$ deg$^2$ across four extragalactic legacy fields: CDFS, COSMOS, ELAIS--S1 and XMM-LSS. The $L$--band receiver is being used for the bulk of the survey work, reaching a nominal sensitivity of $\sim 4\mu$Jy, with additional observations made over a smaller area with the $S$--band receiver. The survey simultaneously provides radio continuum (e.g., \citealp{2022MNRAS.509.2150H, 2022MNRAS.516..245W, 2023MNRAS.520.2668H, 2025MNRAS.536.2187H}), spectral line (e.g., \citealp{2021A&A...646A..35M, 2021MNRAS.508.1195P, 2023MNRAS.522.5308P,2024MNRAS.534...76H}) and polarisation (e.g., \citealp{2023A&A...678A..56B,2024MNRAS.528.2511T}) data, allowing a broad range of science topics to be addressed. 

\subsubsection{Catalogues and photometric flux measurements}

The first step in our analysis is to identify the fraction of our sample (501 DES-SN hosts) that have associated radio observations. We use host galaxy matched catalogues compiled by Hale et al. (in prep), which provide $K_{\rm s}-$band selected counterparts for the radio detected sources in MIGHTEE data release 1 (DR1). The radio sources were associated with the optical/NIR counterpart using a combination of likelihood-ratio technique \citep[e.g.][]{McAlpine2012} and visual inspection to cross-match extended sources.
At present, the catalogues cover three of the four MIGHTEE fields: COSMOS, CDFS and XMM-LSS \citep{2025MNRAS.536.2187H}, but here we only consider the latter two  (see Paper~I for a justification). The DES-SN host galaxy coordinates are positionally crossmatched to the Hale et al. (in prep)  catalogue sources using a 1~arcsec matching radius to the $K_S-$band right ascension and declination, as opposed to the radio counterpart positions, which have a slightly poorer astrometric accuracy due to the lower angular resolution. 
Of the total sample, 95 DES-SN hosts have corresponding radio observations in the Hale et al. (in prep) catalogues.

The Hale et al. (in prep) catalogues only include objects in the MIGHTEE fields that are `detected' in the radio. The standard catalogue selection corresponds to a peak flux density threshold of $5\sigma$ (see \citealp{2025MNRAS.536.2187H} for details). Only considering sources that have significant detections at this wavelength introduces a bias towards studying massive star-forming galaxies (SFGs) and/or active galactic nuclei (AGN). To mitigate this bias and derive a more \textit{complete} sample that includes radio `non-detections', we additionally measure photometry for all DES-SN host galaxies that overlap with the MIGHTEE image footprint in CDFS and XMM.

Radio photometry is performed on MIGHTEE $L-$band images \citep{2025MNRAS.536.2187H}, which are calibrated in units of Jy/beam. MIGHTEE DR1 includes both low- (7.3 and 8.9 arcsec beam FWHM) and high-resolution (5.5 and 5.0 arcsec beam FWHM) images across CDFS and XMM, respectively. We perform photometry on the high-resolution images across both fields. However, in practice, source fluxes at the redshifts of the DES sample ($z < 1.2$) are expected to vary little with image resolution, since the host galaxies are unresolved at both resolutions. We confirm that this is indeed the case. To extract fluxes, we measure the pixel value at the positions of the DES-SN host galaxies, which represent the total flux density of a point source positioned at the centre of that pixel\footnote{The assumption that underlies this is that the sources are unresolved, which holds true for the bulk of our sample.}. The flux values are median background subtracted. The associated errors are derived via a two step process. First, a background distribution is determined by measuring the values in the nearest 500 pixels (an arbitrary choice) to each source position. At these wavelengths, source confusion from multiple unresolved galaxies within a single MeerKAT beam results in a skewed normal background distribution, characterised by an excess of positive source counts. Therefore, the second step in determining the error on the MIGHTEE fluxes is to fit the distribution with an asymmetric Gaussian. We take the right-hand side i.e., larger standard deviation, as our error term (a conservative choice) to account for the issue of source confusion. This yields radio flux measurements for 372 DES-SN galaxies. For this, we use flux measurements from Hale et al. (in prep) where possible (95 galaxies), and our forced-photometry measurements for the remaining.  129 DES SN host galaxies fall outside the currently available data from the MIGHTEE survey and do not bias our results in any way. Any additional selection cuts applied at later stages of the analysis are discussed in the relevant sections.

\subsubsection{1.4 GHz luminosity}

Host galaxy radio flux densities are converted to rest-frame 1.4 GHz radio luminosities using,

\begin{equation} \label{eq:radlum}
    L_\mathrm{\nu_{em}} = \left(\frac{\nu_{\mathrm{obs'}}}{\nu_{\mathrm{em}}}\right)^{\alpha_{\rm rad}} S_\mathrm{\nu_{obs'}} (1+z)^{\alpha_{\rm rad}-1} 4 \pi D_\mathrm{L}^2
\end{equation}

\noindent where $\nu_{\mathrm{em}}$ is the frequency at which the radiation is emitted in the rest frame, $\nu_{\mathrm{obs'}}$ is the frequency at which observations are taken, $\alpha^\text{rad}$ is the radio spectral index, $S_\mathrm{\nu_{obs'}}$ is the flux density and $D_\mathrm{L}$ is the luminosity distance of the source. We take $\nu_{\mathrm{em}} = 1.4$ GHz which is a standard reference frequency in radio continuum astronomy due to the wide availability of $L-$band observations.  
Due to the wide bandwidth of the $L-$band receiver on MeerKAT and the large primary beam (field of view), the effective observed frequency varies across the field, therefore
$\nu_{\mathrm{obs'}}$ is determined by measuring the pixel value at the positions of the DES-SN host galaxies on the MIGHTEE effective frequency maps \citep{2025MNRAS.536.2187H}. We fix the redshift of each galaxy to the spectroscopic host redshift from the DES-SN5YR data release\footref{desdr}. The errors on radio luminosities are propagated from the measured uncertainties on the radio flux density.

\subsubsection{AGN selection cuts}

Radio continuum diagnostics provide a useful tracer of star formation, unbiased by the effects of dust obscuration, but they are not entirely without limitation. In addition to star formation processes, the observed radio output can be generated by active galactic nuclei (AGN). The impact this has on our ability to estimate the SFR is twofold: (i) it contributes to the  scatter in the conversion from radio luminosity to SFR, and (ii) it can lead to an overestimate of the SFR due to the contribution from a central AGN.

In this work we remove potential AGN-like sources in the following way.  The radio luminosity that AGN typically begin to dominate the source counts occurs at around $L_{1.4} > 10^{23.5}$ W Hz$^{-1}$ (e.g., \citealp{2007MNRAS.375..931M,2008MNRAS.388.1335W}). Of our total sample, eight objects are identified above this threshold and are represented by open circles in all figures. Further inspection of the MIGHTEE radio maps shows two more objects that lie below this threshold but display clear morphological signs of AGN activity in the radio maps i.e. jet-like phenomena. These are represented by open stars in all figures and we omit all ten objects from subsequent analyses, yielding a final sample comprised of 362 DES-SN galaxies. 

More sophisticated methods of identifying AGN are often applied to deep radio surveys \citep[e.g.][]{2022MNRAS.516..245W}, but these generally rely on having deep data at other wavelengths, beyond the optical and near-infrared. We therefore, do not use them here as our aim is to emulate what will be feasible with the data available across the wide-field survey of LSST. Although we note, such data, where available, could be used to increase sample purity. 

\section{Applications of 1.4 $\mathrm{GHz}$ diagnostics to type Ia supernova  host studies} \label{sec:umbrella}

\subsection{Introduction} \label{sec:results1}

The physical interpretation of radio continuum emission as a tracer of star formation relies on calibration against independent SFR indicators. Calibrations are usually anchored to FIR emission through the tight radio--infrared correlation (referenced to 1.4 GHz for synchrotron-dominated radio emission e.g., \citealp{1992ARA&A..30..575C,2001ApJ...554..803Y,2003ApJ...586..794B, 2010MNRAS.409...92J}). Independent of FIR-based approaches,  recent work (e.g., \citealp{2017MNRAS.466.2312D,Smith2021, 2024MNRAS.531..708C}) has begun to shift its focus to calibrating the SFR--radio relation using SFRs inferred from full SED modelling. This approach combines FIR constraints with information at other wavelengths and is therefore expected to provide a more representative measure of the total star formation, while also allowing a more robust treatment of the associated uncertainties \citep{2016MNRAS.461..458D}. An illustrative example of this methodology is presented by \cite{2024MNRAS.531..708C}, who derive SFRs from \textsc{proSpect}-based \citep{2020MNRAS.495..905R} SED modelling, with recent activity characterised by the SFR$_\text{burst}$ parameter, averaged over a 100 Myr timescale. Their analysis considers a  volume-limited sample of $\sim$5,500 SFGs, combining 1.4~GHz radio data from the MIGHTEE survey--also used in this work--with ancillary survey data. Following their detailed investigation, we adopt the prescription derived therein:

\begin{equation} \label{eq:cook}
    \frac{\text{SFR}}{\text{M}_\odot\text{yr}^{-1}} = 10^{1.014\pm0.003}\cdot \left(\frac{L_{1.4\hspace{0.2em}\text{GHz}}}{5\times10^{22}\text{WHz}^{-1}}\right)^{0.868\pm0.005}
\end{equation}

We note that this specific calibration choice is not unique, and alternative physically motivated formulations exist (e.g., \citealp{1992ARA&A..30..575C,2003ApJ...586..794B,2003ApJ...599..971H,2015A&A...579A.102B,2017MNRAS.466.2312D}); however, many of these either rely on calorimetric assumptions \citep{1992ARA&A..30..575C} or are traditionally anchored to SFRs inferred from single- or limited wavelength diagnostics \citep{2003ApJ...586..794B}. While \cite{2017MNRAS.466.2312D} adopt a similar SED-based approach, their calibration is derived from a significantly smaller sample of 144 galaxies (cf. {$\sim$5500 in \citealp{2024MNRAS.531..708C}}). 

We do not re-derive a SFR--$L_{1.4}$ calibration in this work as our galaxy sample is defined according to explicit criteria (Section \ref{sec: data}) and does not include all sources within the relevant redshift range (i.e. not volume-complete). This has the potential to introduce selection-driven biases into any inferred relation. 
Moreover, the future applicability of 1.4~GHz SFRs to SN~Ia host galaxy studies will likely rely on existing calibrations rather than those derived from the host samples themselves. The \cite{2024MNRAS.531..708C} calibration therefore provides a suitable external reference for our primary aim--assessing whether radio emission can be used  
to separate galaxies by their star-forming properties in the context of this science.

Although we do not attempt to derive a new calibration here, it is nevertheless informative to examine the distribution of our SN host galaxies relative to the \cite{2024MNRAS.531..708C} relation for consistency. This comparison combines $L_{1.4}$ values measured in this work (Section~\ref{sec: data}) with SED-derived SFRs taken from Paper~I. Briefly, these SFRs are inferred using the \textsc{Bagpipes} SED fitting code \citep{2018MNRAS.480.4379C} together with the updated 2016 \cite{2003MNRAS.344.1000B} stellar population synthesis models and a \cite{2001MNRAS.322..231K} initial mass function (IMF). For the remaining user-specified parameters, we assume a log-normal star formation history and a  \cite{2000ApJ...539..718C} dust attenuation law with a variable slope i.e. the fiducial model in Paper I and we refer the reader there for full details and priors.

\begin{figure*} 
  \begin{subfigure}[b]{0.49\linewidth}
    \centering
    \includegraphics[width=1\linewidth]{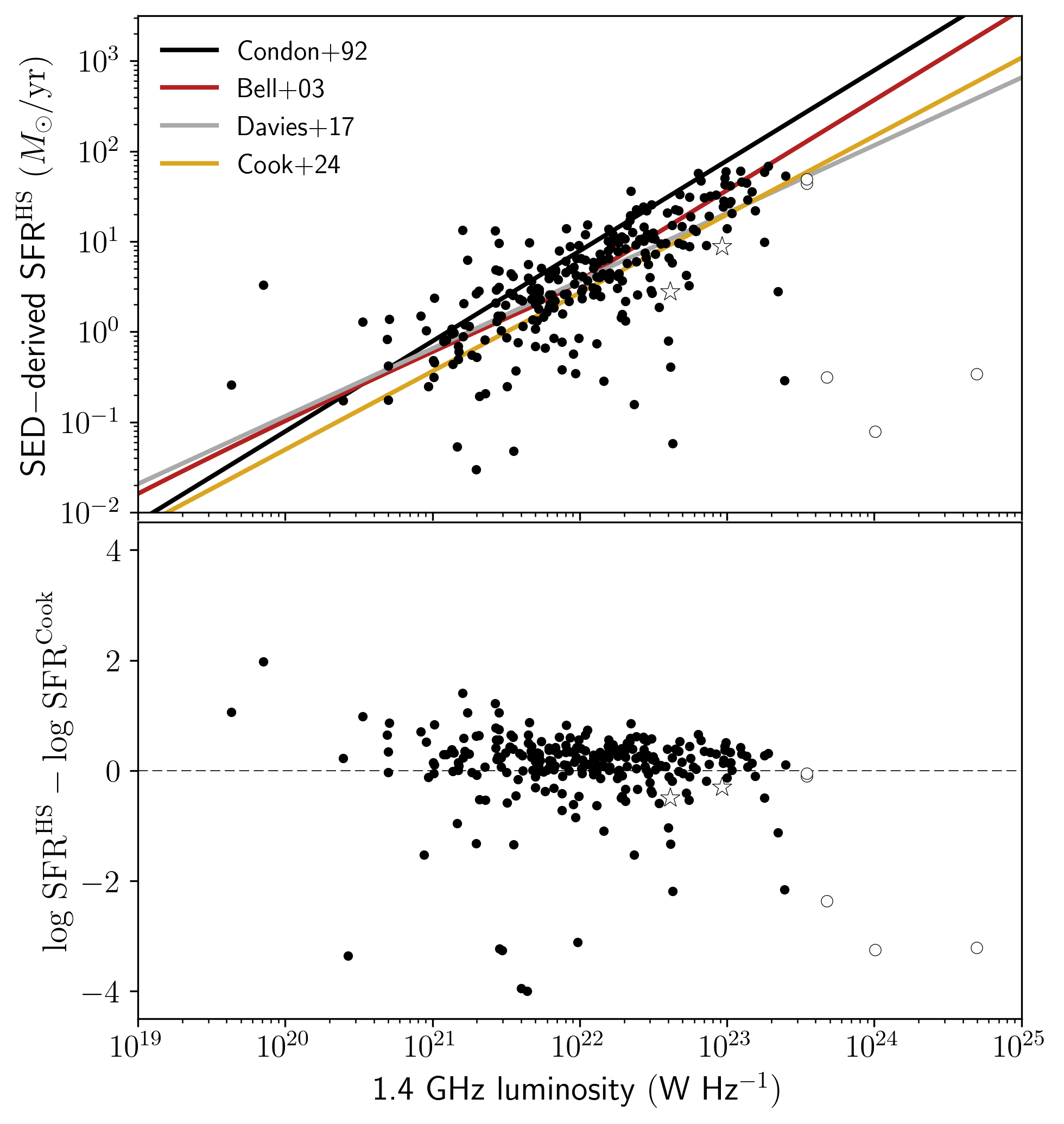} 
  \end{subfigure} \hfill
  \begin{subfigure}[b]{0.49\linewidth}
    \centering
    \includegraphics[width=1\linewidth]{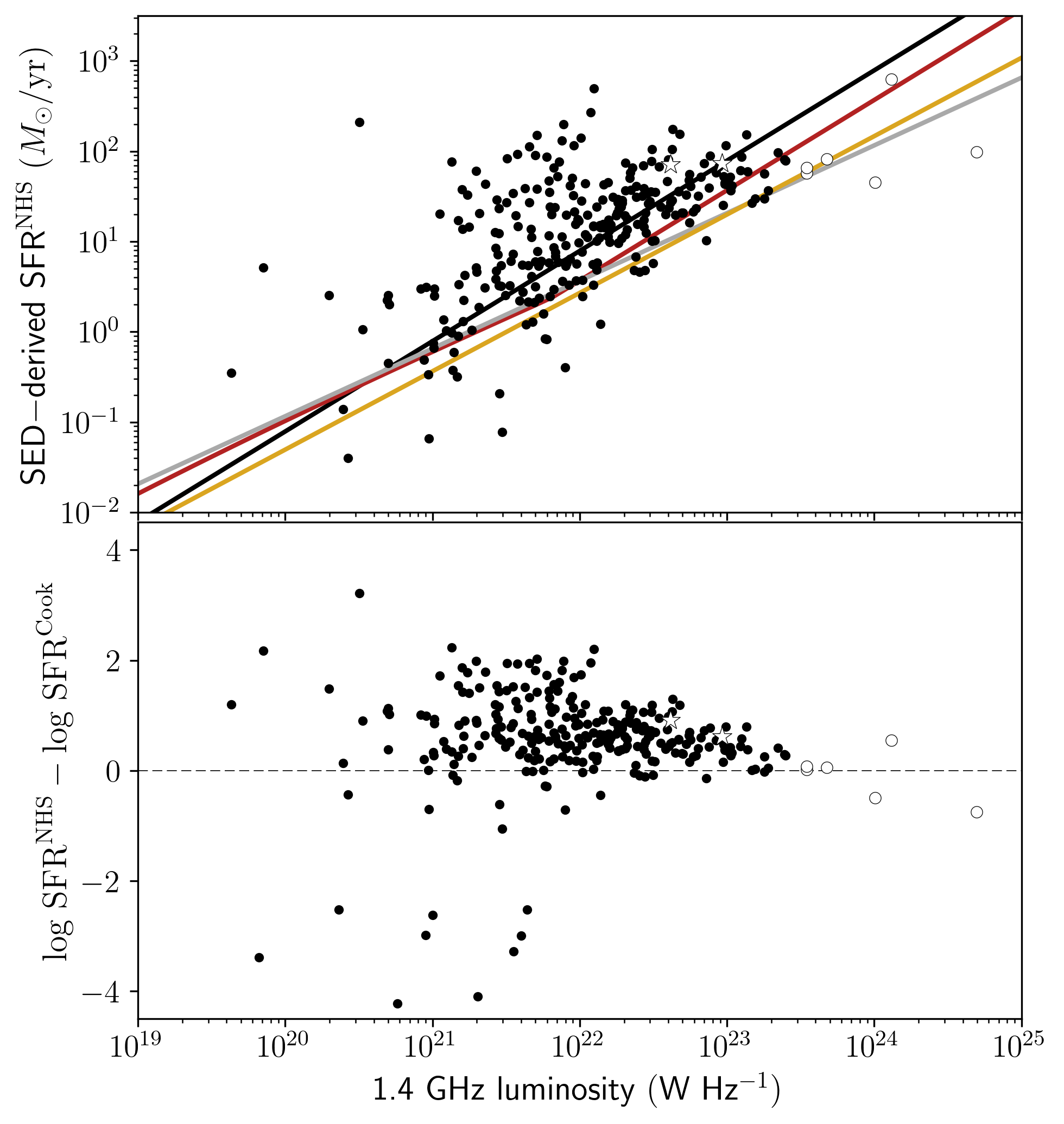} 
  \end{subfigure}
  \caption{Left: SFR$^\mathrm{HS}$ determined from SED fitting of multi-wavelength data (values taken from Paper~I) versus 1.4 GHz radio luminosity, $L_{1.4 \hspace{0.2em}\text{GHz}}$ (computed in this work). The SED-derived estimates are from fitting data that include far-infrared coverage, specifically from the  \textit{Spitzer} and \textit{Herschel} space telescopes (denoted by `HS'; see Paper I). 
  Overplotted are several reference SFR--$L_{1.4 \hspace{0.2em}\text{GHz}}$ calibrations from the literature (\citealp{1992ARA&A..30..575C,2003ApJ...586..794B,2017MNRAS.466.2312D,2024MNRAS.531..708C}) 
  The open circles represent  sources with $L_{1.4} > 10^{23.5}$ W Hz$^{-1}$, i.e. in the AGN-luminosity regime, and the open stars are objects with morphological signs of AGN activity in the radio maps (Section~\ref{sec: data}).
  Error bars are removed for clarity.  Right: as left, except the far-infrared data are excluded when deriving SFR from SED fitting (`NHS' refers to `No \textit{Herschel} and \textit{Spitzer}').} 
  \label{fig:bagvsradio} 
\end{figure*}

Fig.~\ref{fig:bagvsradio} shows the SFR--$L_{1.4}$ relation for our SN host galaxies using SED-derived SFRs from \textsc{Bagpipes}, together with residuals shown relative to the \cite{2024MNRAS.531..708C} calibration (gold solid line). This relation is presented for two SED fitting configurations: the left panels use SFRs derived from fits to the full multiwavelength dataset available in Paper~I, including \textit{Herschel} and \textit{Spitzer} (`HS') measurements, while the right panel shows the same relation for SFRs derived without these FIR data (`NHS' stands for `No \textit{Herschel} and \textit{Spitzer}'). This HS/NHS notation is used consistently for other SED-derived quantities throughout the paper. 

For clarity, the axes are truncated and shown in logarithmic scale; sources lying outside the plotted range, as well as objects with negative 1.4~GHz flux densities, are not displayed. 
Error bars are also removed for clarity. As Fig.~\ref{fig:bagvsradio} is used solely to illustrate the relative distributions of galaxies in the HS versus NHS cases, these visualisation choices do not affect any subsequent results or analysis, which makes use of the full galaxy sample irrespective of whether individual sources appear in this figure specifically.

Several other literature calibrations from \cite{1992ARA&A..30..575C,2003ApJ...586..794B} and \cite{2017MNRAS.466.2312D} are overplotted for reference. For consistency with Paper~I, all relations are scaled to a \cite{2001MNRAS.322..231K} IMF, using conversions outlined in  \cite{2014ARA&A..52..415M}, for \cite{1955ApJ...121..161S} and \cite{2003PASP..115..763C}, and \cite{2000ApJ...544..641H} for \cite{1979ApJS...41..513M}.

The HS and NHS configurations are shown to illustrate the sensitivity of the inferred SFRs--and hence the appearance of the SFR--$L_{1.4}$ relation--to the available wavelength coverage used in SED fitting. In the HS configuration (left panels of Fig.~\ref{fig:bagvsradio}), the SED-derived SFRs show broad agreement with the \cite{2024MNRAS.531..708C} relation (equation~\ref{eq:cook}) across the sampled range in $L_{1.4}$, with residuals ($\Delta\rm SFR =\rm SFR_{\rm SED} - SFR_{\rm Cook}$) approximately centred on zero,  having a median of 0.42  $\rm M_\odot/ yr$, a 16$^{\rm th}$-84$^{\rm th}$ percentile range of [-1.72, 6.12] and an r.m.s scatter, $\sigma_{\Delta\rm SFR} = 8.61\pm0.64 \hspace{0.2em} \rm M_\odot/\rm yr$. In the absence of FIR data (right panels of Fig.~\ref{fig:bagvsradio}), the SFR estimates are subject to stronger degeneracies with dust attenuation and stellar population age, allowing a wider range of star formation histories to reproduce the observed optical--NIR photometry. This manifests as the inferred SFRs displaying markedly increased scatter and a systematic offset toward higher values relative to the \cite{2024MNRAS.531..708C} relation (see Paper~I, Section~6), with a broader residual distribution, characterised by a median of 8.48 $\rm M_\odot/yr$, a 16$^{\rm th}$-84$^{\rm th}$ percentile range of [-0.10, 37.7] and an r.m.s scatter, $\sigma_{\rm \Delta SFR} =44.3\pm3.30 \hspace{0.2em} \rm M_\odot/yr$. Taken together, this comparison underscores how SED-fitting choices can propagate into the radio--SFR relation, with direct implications for studies that seek to calibrate or interpret them.

\subsection{1.4 GHz SFR -- \textit{M}$_*$ plane} \label{sec:results2}

\begin{figure*}
    \centering
\includegraphics[width= 170mm]{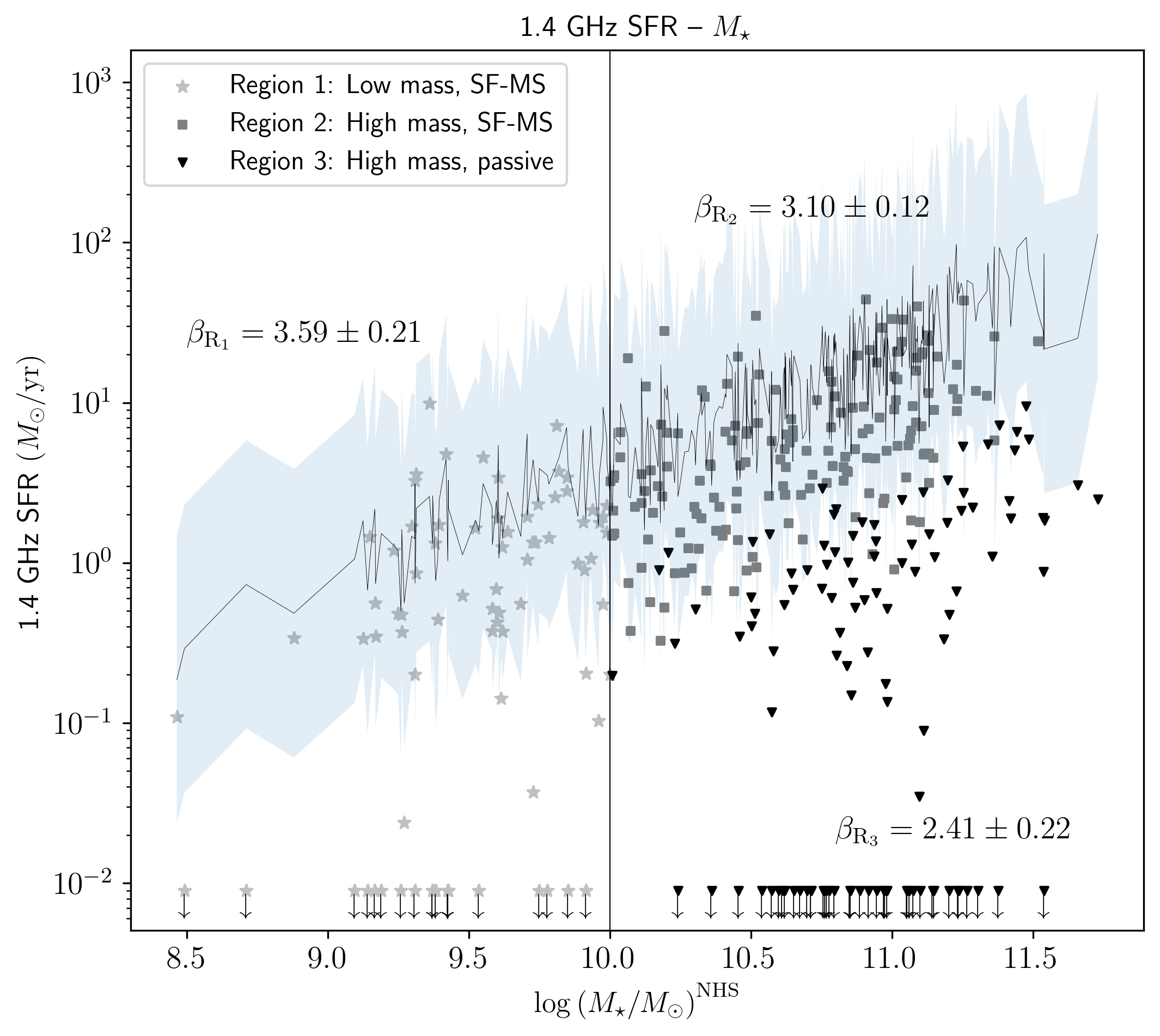}
    \caption{The distribution of SN Ia host galaxies across the 1.4 GHz SFR--$M_*$ plane. The 1.4 GHz SFRs are computed directly in this work, whereas the galaxy stellar masses, $M_*$ are determined and taken from Paper~I. The plane is divided into three regions (coded by marker type) to isolate galaxies that are characterised by similar properties. Low-mass galaxies are indicated by stars, where we apply an arbitrary cut on mass at 10$^{10}\rm M_\odot$. High-mass galaxies (> 10$^{10}\rm M_\odot$) are then split into two populations based on whether they lie within a 3$\sigma$ interval of the SF-MS, which is highlighted by the blue shaded region. High-mass, high SFR galaxies are shown by squares and conversely, high-mass, low SFR (passive) galaxies are shown as triangles. Galaxies with SFR values below 0.01 $\rm M_\odot/$yr are set as upper limits to this value and are indicated by downward arrows for illustration purposes. Refer to Table \ref{beta-vals} for a full summary of cosmological SN Ia nuisance parameter constraints.}
    \label{fig:radio_SF-MS}
\end{figure*}

Using the SFR$-L_{1.4}$ calibration from \cite{2024MNRAS.531..708C}, we construct the 1.4 GHz SFR--$M_\star$ plane (Fig. \ref{fig:radio_SF-MS}) for our sample of DES-SN hosts. This Figure is intended as a replica of Fig.~7 in Paper~I, but with 1.4 GHz SFRs in place of \textsc{Bagpipes} SED-based SFRs. The stellar masses are still computed from SED-fitting and are taken from Paper~I, with one exception; \textit{Herschel} and \textit{Spitzer} data are excluded from their computation. This is to assess the reproducibility of the SFR--$M_\star$ relation presented in Paper~I for LSST sources lacking mid-/FIR coverage.

Next, we split the 1.4 GHz SFR--$M_\star$ plane into three regions. The first region comprises SNe~Ia in low-mass hosts, where we define the threshold between low- and high-mass hosts at $10^{10}\rm M_\odot$. To split high-mass hosts into star-forming versus passive systems, we adopt the parametric form of the SF-MS presented in \cite{2012ApJ...754L..29W}. While alternative parametrisations exist \citep[e.g.,][]{2014ApJS..214...15S, 2014ApJ...795..104W,2015A&A...575A..74S}, we choose the \cite{2012ApJ...754L..29W} relation to maintain internal consistency with Paper~I, ensuring that any regional differences can be attributed solely to the change in SFR tracer, rather than to shifts in the classification scheme. In its mathematical form, the SF-MS is expressed as,

\begin{equation}\label{eq:whit}
    \text{log}_{10}[\text{SFR(\textit{z})}] = \alpha (z)[\text{log}_{10}(M_\star) - 10.5] + \beta(z) 
\end{equation}

\noindent where $\alpha(z)$ and $\beta(z)$\footnote{These $\alpha$ and $\beta$ parameters are not to be confused with those used to characterize the SN stretch- and colour-luminosity relations.} parametrise the slope and normalisation of the SF-MS, respectively and are given by,

\begin{equation*} 
\begin{array}{c}
    \alpha(z) = \alpha_1 + \alpha_2z \\[1em]
    \beta(z) = \beta_1 + \beta_2z + \beta_3z^2
\end{array}
\end{equation*}

\noindent with fit parameters: $\alpha_1$, $\alpha_2$, $\beta_1$, $\beta_2$ and $\beta_3$. We take the median values presented in Table 3 in \cite{2015MNRAS.453.2540J}. Adopting this parameterisation, the second region isolates galaxies that lie within a 3$\sigma$ interval of the SF-MS (this applies only to galaxies below the SF-MS, not above it). We take $\sigma = 0.3$ dex following \cite{2015MNRAS.453.2540J} (hence $3\sigma = 0.9$ dex). The third and final region, consists of passive galaxies that have evolved off and lie $>3\sigma$ below the SF-MS. Full details are presented in Paper I.

\subsubsection{Comparison to Paper I regions}

\begin{figure*} 
  \begin{subfigure}[b]{0.49\linewidth}
    \centering
    \includegraphics[width=1\linewidth]{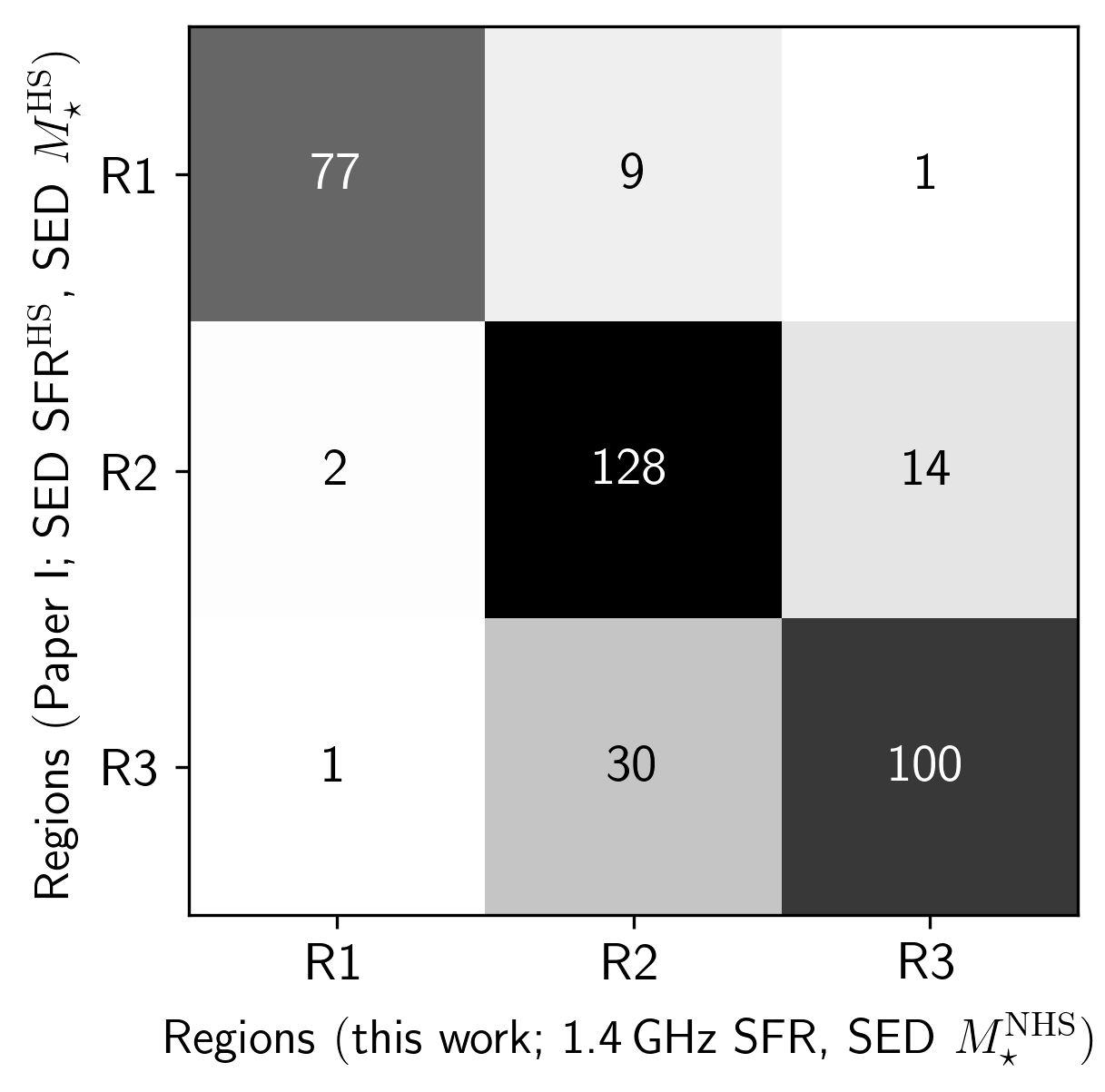} 
  \end{subfigure} \hfill
  \begin{subfigure}[b]{0.49\linewidth}
    \centering
    \includegraphics[width=1\linewidth]{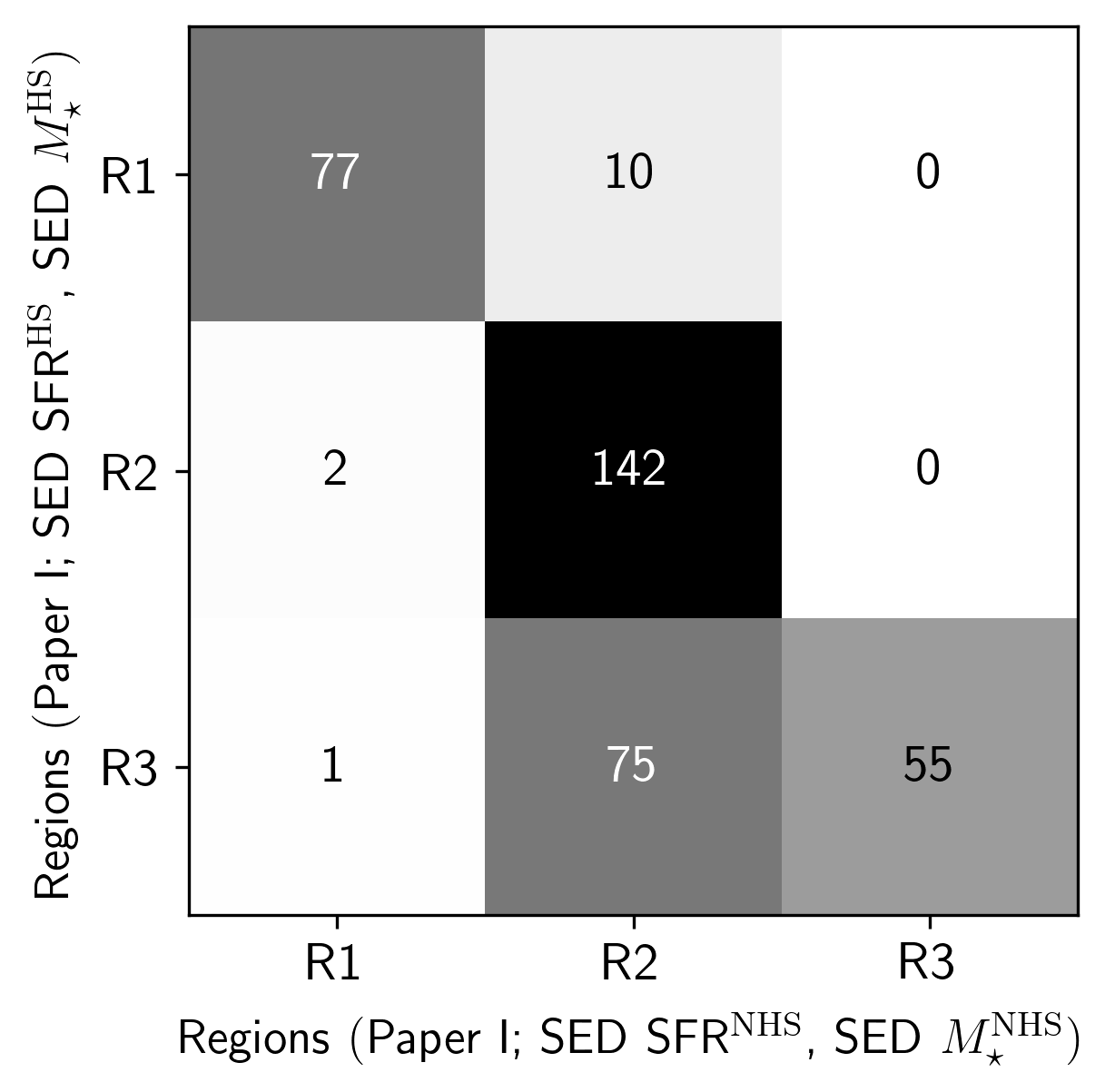} 
  \end{subfigure}
  \caption{Left: Confusion matrix comparing region classifications from Paper I, based on SED-derived galaxy properties including \textit{Herschel} and \textit{Spitzer} (`HS') data, with region classifications from this work, based on 1.4 GHz-derived SFRs (computed using the \protect\cite{2024MNRAS.531..708C} relation) and SED-derived stellar masses excluding \textit{Herschel} and \textit{Spitzer} (`NHS') data. The values in each cell give the number of galaxies from a given Paper I region (rows) that are assigned to the corresponding region in this work (columns). Right: Same as left, but comparing Paper I region classifications based on SED-derived galaxy properties including (`HS') and excluding (`NHS') \textit{Herschel} and \textit{Spitzer} data.} 
  \label{fig:confusionm} 
\end{figure*}

The three region subsamples (low-mass SF, high-mass SF and high-mass passive) from this work can now be compared to those from Paper~I using a confusion matrix to assess the level of consistency between the two classification schemes (Fig. \ref{fig:confusionm}, left panel). The Paper~I region assignments, based on SED-derived SFRs and stellar masses including \textit{Herschel} and \textit{Spitzer} (`HS'), are treated as the reference ($y-$axis). Importantly, the  classification shown on the $y-$axis should not be interpreted as an absolute ground truth, but rather as a higher-information reference against which we assess the consistency of alternative selection methods. Equivalent confusion matrices constructed using radio--SFR calibrations alternative to \cite{2024MNRAS.531..708C} are presented in Appendix~\ref{app:alt_radsfr}.

The resulting confusion matrix is strongly dominated by the diagonal, with  ${\sim} 84$ per cent of DES-SN hosts retaining the same regional classification between this work and  Paper~I. To interpret the consistency of individual regions, we focus on the purity of the radio-selected subsamples (i.e. interpreting the matrix column-wise), rather than their completeness with respect to the reference classification. This distinction is particularly relevant for LSST-era applications, where the primary requirement is the construction of clean SN host galaxy populations rather than the recovery of all systems identified in higher-information analyses. 

Under this interpretation, the radio-selected low-mass sample (Region~1) shows minimal contamination, with  ${\sim} 96$ per cent of galaxies in this category corresponding to the same region in the reference classification. This demonstrates that the low-mass region remains cleanly isolated on the SFR--$M_\star$ plane despite the removal of far-infrared measurements in the SED fitting.

For the high-mass populations (Regions~2 and 3), the agreement between the two classification schemes remains strong, though a higher degree of mixing is observed compared to the low-mass regime; expectedly, given that these regions are separated primarily by their star-formation properties and are therefore sensitive to the choice of SFR tracer. The radio-selected Region~2 sample exhibits a purity of ${\sim}77$ per cent, with the dominant contamination arising from galaxies assigned to the passive region in the reference classification (30 sources). Conversely, the radio-selected Region~3 sample exhibits a purity of ${\sim}87$ per cent, with most of the contamination arising from Region~2 in the reference sample (14 sources). 

Some fraction of the off-diagonal population may arise from differences in tracer timescales, particularly in systems with non-steady star formation histories (SFHs). \cite{2024MNRAS.531..708C} find that galaxies with increasing SFHs can exhibit elevated SED-derived SFRs relative to their radio emission, which they attribute to the finite time required for supernova-driven radio emission to build up. Complementarily, \cite{2023A&A...675A.126A} show that galaxies with declining SFHs can exhibit radio-derived SFRs in excess of their instantaneous SED-derived SFRs, with roughly $30\%$ of their sample showing radio SFRs at least an order of magnitude higher, consistent with radio emission responding to changes in star formation over longer timescales ($\sim100$~Myr). More generally, this behaviour may arise from the choice of adopted calibration or from systems lying near the star-forming--passive boundary. In addition to this overall scatter, the mixing appears asymmetric in nature, with a larger number of galaxies \textit{misclassified} relative to the reference axis from the passive (Region~3) to the star-forming (Region~2) sample than vice versa. This may partly reflect low-level AGN activity, which can enhance radio emission without a corresponding increase in SED-based SFR estimates.

For completeness, Fig. \ref{fig:confusionm} (right panel) shows a confusion matrix constructed in the same manner as the left panel, but with both SFRs and stellar masses derived from SED fitting excluding \textit{Herschel} and \textit{Spitzer} data ($x-$axis), while retaining the same reference ($y-$axis). The percentage of SN hosts assigned identically (i.e. along the diagonal) between the SED$^\text{HS}$ and SED$^\text{NHS}$ cases drops to $76$ per cent.

When read column-wise, the Region~1 classifications in this matrix are unchanged relative to the left panel, as the SED-derived stellar masses in both matrices are derived without far-infrared data. The more informative comparison lies in the high-mass regime (Regions~2 and 3), where the separation is driven primarily by star-formation activity. For the SED$^{\rm NHS}$-selected Region~2 sample, substantial mixing is shown with the reference passive (Region~3) population, with a purity of $62$ per cent. By contrast, the SED$^{\rm NHS}$-selected Region~3 sample consists of a much narrower subset of the passive population identified in Paper~I, recovering fewer than half of the galaxies compared to the radio-based selection (left panel matrix), despite exhibiting minimal contamination. Such a reduced, contamination-free sample should not necessarily be interpreted as a cleaner representation of the same population, as it may instead preferentially select galaxies with particular SED characteristics in the absence of far-infrared constraints, thereby potentially introducing unquantifiable selection effects.

Quantifying the origin of this behaviour is not the focus of this work and we include the NHS case only for comparison, as many previous SN host studies rely on similarly limited wavelength coverage. We do not explore or comment on this further. Instead, our emphasis is on the radio-based classification, the use of which is more physically motivated \citep{1992ARA&A..30..575C} and thus, forms the primary basis of our analysis. The strong diagonal dominance and high regional purities in the confusion matrix (Fig.~\ref{fig:confusionm}, left panel) demonstrate that 1.4 GHz SFRs provide a robust and internally consistent alternative to FIR-based SED SFRs, preserving a broader and more representative high-mass population while maintaining a clear separation between regions on the SFR--$M_\star$ plane. However, as these inferences are drawn from the confusion matrix alone, we next assess their robustness by examining whether SN~Ia subsamples based on the radio-selected regions exhibit consistent properties with those measured in Paper~I.

\subsubsection{Nuisance parameter constraints}

In this section we investigate the SN nuisance parameters ($\alpha,\beta$)
that are used to correct for variations in SN luminosities, as well as the absolute magnitude of SNe Ia ($M$ in equation \ref{eq:distmod}). We note, however, that without an absolute calibrated distance scale, $M$ is degenerate with $H_0$. Instead, cosmological fits typically marginalise over a single parameter, $\mathcal{M} = M + \text{log}_{10}(c/H_0)+25$, that combines these terms. Therefore, we do not present our results in terms of $M$, which requires an assumption for the value of $H_0$ but choose to report out findings in terms of $\Delta \mathcal{M}$, the relative value of $\mathcal{M}$ with respect to our full sample constraint.  Full details of the procedures used in the computation of the parameters are presented in Paper I but, briefly, they are measured using the ``BEAMS with Bias Corrections'' (BBC; \citealp{2017ApJ...836...56K}) framework. We obtain constraints both for our full sample and for the subsamples of SNe Ia in each division of the 1.4 GHz SFR-$M_*$ plane. Our results are presented in Table \ref{beta-vals} and findings are as follows,

\begin{enumerate}[label=(\roman*), leftmargin=*,align=left,labelindent=\parindent, labelsep=0.5em, itemindent=!]

\item The values of $\alpha$ in each of the three sub-regions are consistent, with the largest difference at the ${\sim}1.9\sigma$ level (between regions 2 and 3).

\item SNe Ia in low-mass hosts (Region~1) are characterised by higher values of $\beta$ than those in high-mass hosts. The largest difference (${\sim}3.9\sigma$) exists between regions 1 and 3 (passive hosts). Smaller differences in $\beta$ are seen between regions 1 and 2 (high mass, MS), significant at the ${\sim}2.0\sigma$ level. Differences persist between high-mass hosts (regions 2 and 3), with passive hosts displaying the smallest $\beta$ values (${\sim}2.8\sigma$).

\item Consistent with previous works \citep{2010ApJ...715..743K,2010MNRAS.406..782S,2010ApJ...722..566L}, we find SNe~Ia in high-mass hosts (regions 2 and 3) are, on average, brighter (negative $\Delta\mathcal{M}$) post-standardisation than those in low-mass hosts. This well known trend is referred to as the `mass step' and has been confirmed in previous cosmological analyses that fit the full SN~Ia sample. The mass step is usually defined as the difference between the average Hubble residuals for low- and high-mass galaxies. As we are minimising the residuals for each region separately, we instead look at differences in the SN absolute rest-frame magnitude across the three regions. The most significant brightness differences (0.095 mag) are seen between SNe~Ia in low-mass (Region~1) and high-mass passive (Region~3) hosts, at a ${\sim}3.0\sigma$ significance level. The smallest differences in brightness (${\sim}0.032$ mag) are between SNe in both high mass regions (${\sim}1.2\sigma$). Differences between low-mass and high-mass MS (Region~2) exist at the ${\sim}2\sigma$ significance level. 

\end{enumerate}

While the qualitative trends observed here are similar to those reported in Paper~I, the statistical significance of several inter-region differences changes when using radio-derived SFRs. This is expected, as the classification of host galaxies depends on the tracer used to characterise star formation activity, and the resulting redistribution of SNe between regions naturally alters the measured contrasts in the nuisance parameters.

Nevertheless, the nuisance parameters inferred in each region remain consistent with those obtained using SED-based SFR classifications at the ${\sim}1.1\sigma$ level (see Table~\ref{beta-vals}). Region~3 provides a particularly stringent test of the impact of SFR estimator choice on the inferred SN parameters. In Paper~I, we demonstrate that the inferred value of $\beta$ in Region~3 is particularly sensitive to contamination from SNe Ia associated with neighbouring star-forming galaxies, with the constraint responding systematically to shifts in the boundary between Regions~2 and 3 (see Paper I,  Fig.~9). The fact that the radio-SFR based analysis reproduces the Paper~I results within the uncertainties indicates that radio SFRs offer a promising basis for partitioning galaxies in this parameter space, even in regimes where cross-region contamination effects are expected to be most pronounced. 

We note that the Paper~I values reported in Table~\ref{beta-vals} are derived from the primary 501-host `gold' sample; repeating the SED-based Paper~I analysis on the restricted 362-galaxy subset used here yields consistent nuisance parameter constraints.

Additionally, we verify that the stretch and colour distributions of SNe~Ia in each of the three regions are consistent with those reported previously (see Paper~I, Fig.~10).

\begin{table*}
    \centering
    \caption{Nuisance parameter values constrained for SNe Ia in host galaxies that occupy three distinct regions of the SFR--$M_\star$ plane for this work (based on 1.4~GHz SFRs) and the Paper~I results (based on SED-derived SFRs), which are included solely for comparison.}
    \label{beta-vals}
	\begin{tabular}{l c c c c c c c c }  
		\hline
		\hline
         &  &  & This work  &  &  & & Paper I   & \\
        \hline
		\textbf{SF-MS Region} & $N_\mathrm{SN}$$^{(*)}$ & \textbf{$\alpha$} &\textbf{$\beta$} & $\Delta \mathcal{M}$$^{(\dag)}$ &  $N_\mathrm{SN}$ & \textbf{$\alpha$} & \textbf{$\beta$} & $\Delta \mathcal{M}$ \\
        \hline
        \textbf{Full sample} & 358 & 0.162$\pm$0.011 & 3.14$\pm$0.09 & -- &  495 & 0.157$\pm$0.009 & 3.11$\pm$0.08 &  -- \\
        \hline
        \textbf{Region 1:} Low-mass, SF-MS  & 77 & 0.170$\pm$0.030 & 3.59$\pm$0.21 & 0.053$\pm$0.025 &  116 & 0.185$\pm$0.023 & 3.51$\pm$0.16 & 0.055$\pm$0.019\\
        \hline
        \textbf{Region 2:} High-mass, SF-MS & 166 & 0.199$\pm$0.020 & 3.10$\pm$0.12 & -0.010$\pm$0.017 &  197 & 0.180$\pm$0.019 &3.15$\pm$0.11 & 0.000$\pm$0.015 \\
        \hline
        \textbf{Region 3:} High-mass, passive & 115 & 0.147$\pm$0.019 & 2.41$\pm$0.22 & -0.042$\pm$0.020 &  182 & 0.172$\pm$0.013 & 2.12$\pm$0.16 & -0.066$\pm$0.014\\
        \hline
    \end{tabular}
         \vspace{2mm}

\begin{minipage}{\textwidth}
\raggedright
\footnotesize
$^{(*)}$ The number of SNe Ia included in the BBC fit include a $4\sigma$ outlier rejection. For this reason, the BBC fit of the full sample for this work (362 SNe), for example, only includes 358 SNe.  \\
$^{(\dag)}$ $\Delta \mathcal{M}$ is the difference in absolute brightness compared to the full `gold' sample defined in this work, with the errors given in quadrature. 
The quoted uncertainties on $\Delta \mathcal{M}$ should be regarded as a conservative upper bound as the assumption of statistical independence -- which underlies the quadrature addition of uncertainties -- is not strictly valid in this case (each region sample is a subset of the gold sample which means the error on $\mathcal{M}$ for the gold sample and any region subsample are correlated). The true uncertainty on $\Delta \mathcal{M}$ is expected to be lower. \\
\end{minipage}
\end{table*}

\section{Predictions} \label{sec:predictions}

A key question is whether current and forthcoming radio continuum surveys are sufficiently deep to distinguish star-forming from passive  galaxies relative to the evolving SF-MS. This can be evaluated  by converting the detection limit of a given radio survey into a corresponding star-formation rate threshold and comparing it with the evolving SF-MS as a function of redshift. Ambiguity arises when a radio non-detection may correspond to either a passive galaxy or a SFG below the survey sensitivity (complicating a robust separation between Regions~2 and 3). Formally, the condition for \textit{complete} detection of SFGs can be generalised to,

\begin{equation}\label{eq:thresh}
    \text{SFR}_\text{lim}(z) \leq \text{SFR}_\text{MS}(M_\star = 10^{10}\rm{M}_\odot,\it{z}) - \rm 0.9 \hspace{0.2em}\rm dex,
\end{equation}

\noindent where \text{SFR}$_\text{lim}(z)$ denotes the SFR corresponding to the survey radio detection threshold at redshift $z$, derived from the radio--SFR calibration of \citet[equation~\ref{eq:cook}]{2024MNRAS.531..708C}, and SFR$_\text{MS}$ is the SFR of a galaxy of stellar mass $M_\star$ on the SF-MS at the same redshift, for which we take the SF-MS parametrization of \citet[equation~\ref{eq:whit}]{2012ApJ...754L..29W}. The offset of 0.9~dex below the ridge line reflects our adopted definition of the lower boundary of the star-forming population, i.e. galaxies above this threshold are classified as SFGs (Region~2) and those below it as passive (Region~3). This condition is applied only to galaxies with $\geq 10^{10}\rm M_\odot$; systems below this threshold are assigned to Region~1 on the basis of stellar mass alone (constraints from optical/NIR surveys). 

In what follows, we consider the redshift range over which the bulk of cosmologically useful SNe~Ia are expected to be observed--focusing in practice on LSST--and examine how the depths of the overlapping radio surveys compare to the threshold defined by equation~\ref{eq:thresh}, and thus what they imply for the expected detectability and classification of SN~Ia host galaxies.

We emphasise that equation~\ref{eq:thresh} provides a simplified representation, as it relies on the choice of SF-MS parametrization and radio--SFR calibration as proxies for the underlying galaxy population, neither of which are \textit{absolute}. Intrinsic scatter and various systematics are not explicitly accounted for, and the resulting limits on detectable SFR as a function of redshift should therefore be interpreted as approximate forecasts rather than exact thresholds.

\subsection{Deep-field strategy: LSST DDFs--MIGHTEE}

Radio coverage of the LSST Deep Drilling Fields (DDFs) is provided by the ongoing MIGHTEE survey \citep{2016mks..confE...6J}, which will ultimately cover ${\sim}20$ deg$^2$. Over the DDFs, the bulk of cosmologically-useful SNe~Ia from LSST are expected to lie at $z\lesssim 1$ \citep{2023ApJ...944..212M}. 
Taking $z=1$ as the upper bound of the relevant range,  we evaluate equation~\ref{eq:thresh} using the current MIGHTEE survey rms sensitivity of ${\sim}4\mu$Jy ($1\sigma$). The corresponding $3\sigma$ and $5\sigma$ detection thresholds translate into the dashed horizontal SFR limits shown in the top panel of Fig.~\ref{fig:mocksfms}, overlaid on the SF-MS and its adopted 0.9~dex envelope at this redshift. At $z=1$, a galaxy with stellar mass $M_{\star} = 10^{10}\rm M_\odot$ lying on the SF-MS would be expected to have SFR$\sim\rm 10\,M_\odot/$yr, comparable to the MIGHTEE 3$\sigma$ depth. It would therefore be difficult to robustly place such a galaxy into Region~2 or 3, under our adopted definition of the star-forming locus (shaded locus in Fig.~\ref{fig:mocksfms}). 

However, since optical/NIR imaging from SN surveys will provide precise host galaxy positions a priori, forced photometry can be performed (as done in this work) which would allow for flux measurements below the nominal radio $3\sigma$ detection threshold. The depths at each redshift derived here should therefore be interpreted as conservative limits for direct (blind catalogue-level) detections, while deeper statistical constraints are achievable with positionally informed extraction techniques, for example, fully Bayesian source extraction based on the optical/NIR galaxy positions (e.g., \citealp{Malefahlo2026}).

For stellar masses $\sim 10^{11.5} \hspace{0.2em}\rm M_\odot$, the SFR increases by a factor of 10 to $M_{\star}= 10^{10}\,\rm M_\odot$, such that the full star-forming region lies above the $3\sigma$ radio limit; in this regime, a radio non-detection would reliably indicate a passive (Region~3) galaxy (though galaxies above this mass threshold comprise a relatively small fraction of the population).


\subsection{Wide-field strategy: LSST WFD--ASKAP EMU}

For the LSST Wide-Fast-Deep (WFD) survey, the majority of SNe~Ia relevant for cosmological analyses are expected at $z\lesssim0.5$ \citep[see SN~Ia redshift distribution expected from 4MOST TiDES][figures 10 and 11]{2025ApJ...992..158F}. This limit is mainly set by the requirement that a spectroscopic redshift be obtained for the SN host galaxy, as is standard practice in SN cosmological analyses. Across this footprint, radio observations are available from the ASKAP-EMU \citep{EMU,EMU2025}, reaching an rms sensitivity of $\sim10\mu$Jy. At $z=0.5$, the corresponding $3\sigma$ and $5\sigma$ SFR limits are shown as dashed horizontal lines in the middle panel of Fig.~\ref{fig:mocksfms}. 

\subsection{Future prospects}

Further into the future the Square Kilometre Array Observatory (SKAO) has the potential to reach  depths comparable to those of MIGHTEE, but over thousands of square degrees \citep{2015aska.confE..67P}. This combination of depth and area will enable radio-based star-formation rate measurements for SN host galaxies across the wide-area footprints of future transient surveys, including those conducted with Roman, whose SN~Ia sample is expected to have a median redshift of $z\sim1$ and extend to $z\sim2.5$ \citep[Fig.~6]{2025ApJ...988...65R}. 

\begin{figure}
  \centering

  \begin{subfigure}[b]{0.91\linewidth}
    \centering
    \includegraphics[width=\linewidth]{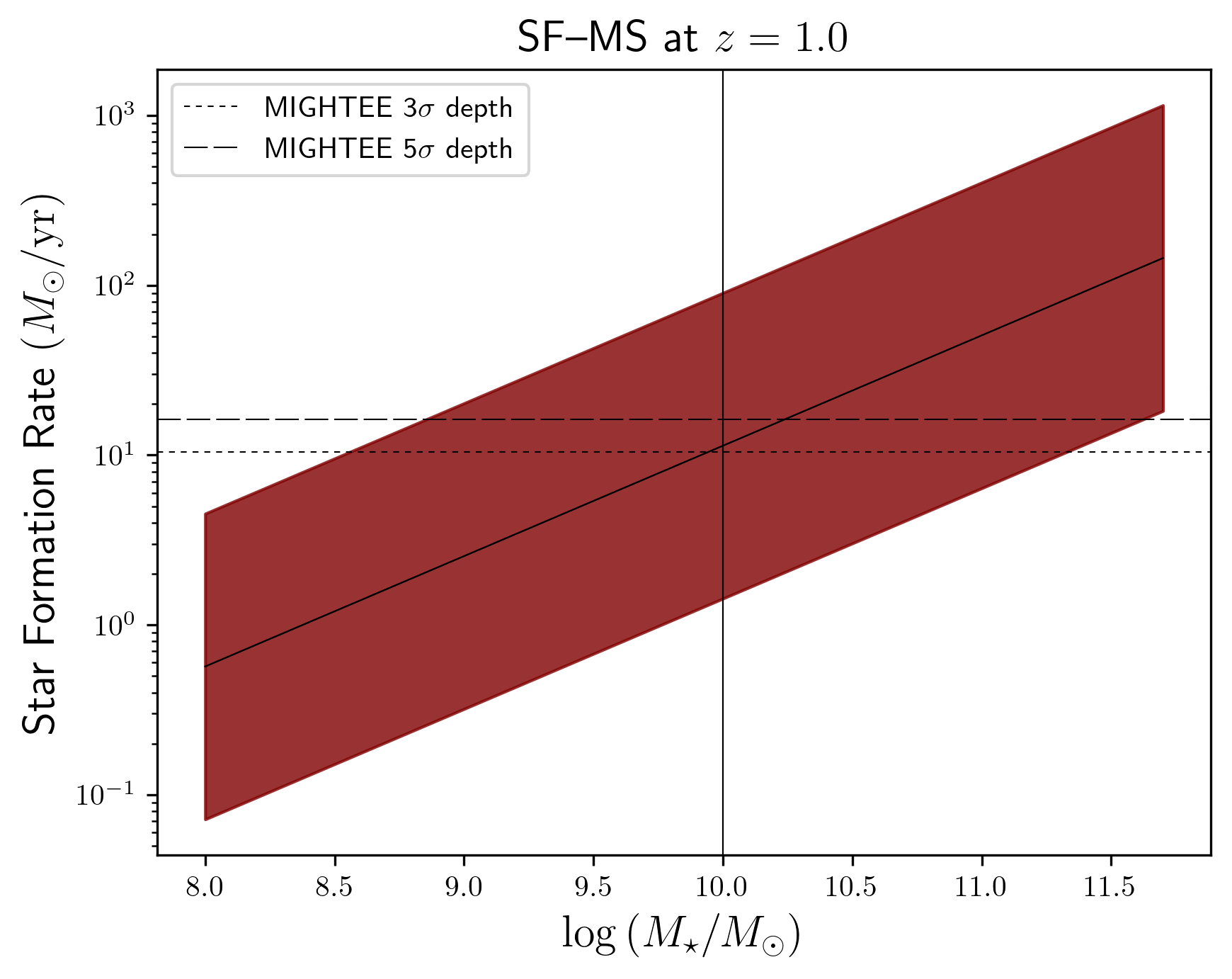}
    \caption{LSST DDFs with MeerKAT/MIGHTEE (rms = $4\mu$Jy) at $z=1$. 
    }
    \label{fig:mightee}
  \end{subfigure}

  \vskip\baselineskip

  \begin{subfigure}[b]{0.91\linewidth}
    \centering
    \includegraphics[width=\linewidth]{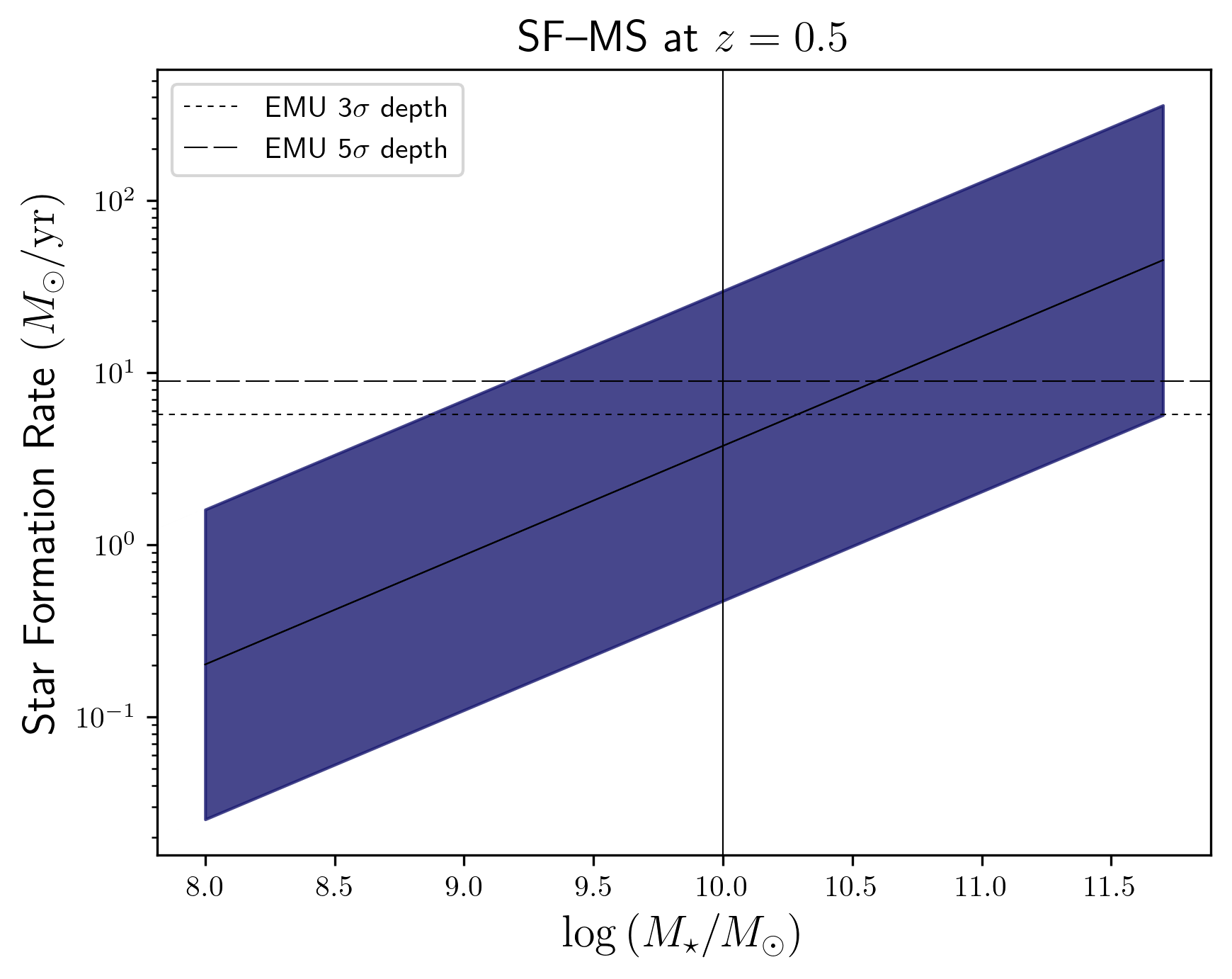}
    \caption{LSST WFD with ASKAP/EMU (rms = $10\mu$Jy) at $z=0.5$. 
    }
    \label{fig:emu}
  \end{subfigure}

  \vskip\baselineskip

  \begin{subfigure}[b]{0.91\linewidth}
    \centering
    \includegraphics[width=\linewidth]{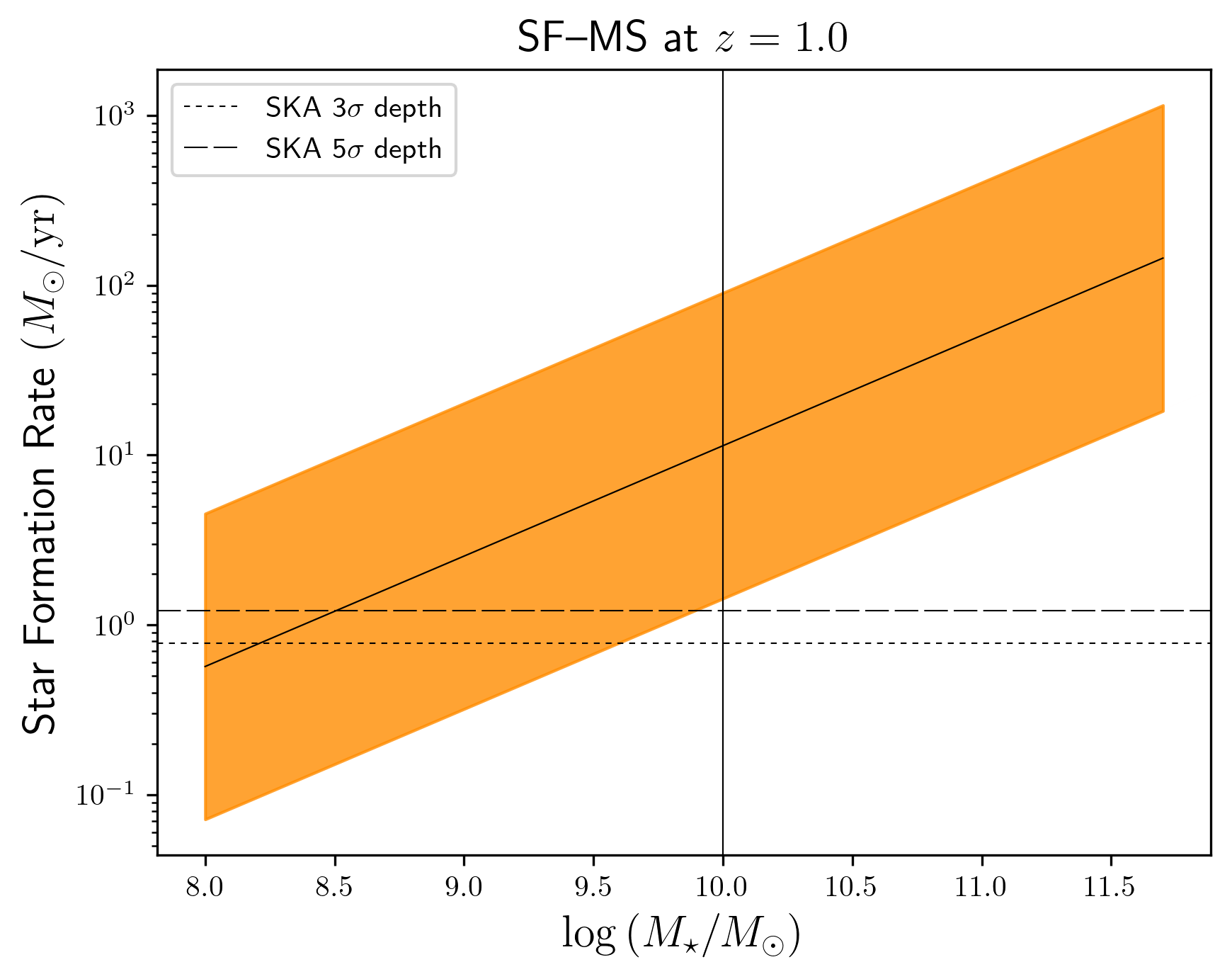}
    \caption{Prospective SKAO deep fields (rms = $0.2\mu$Jy) overlapping LSST DDFs and future Roman SN fields at $z=1$.
    }
    \label{fig:ska}
  \end{subfigure}

  \caption{SF-MS compared to radio detection thresholds at redshifts relevant for SN cosmology. The solid line shows the adopted SF–MS parametrization at the indicated redshift, while the shaded region denotes the star-forming population (Region~2) defined as $\pm$0.9 dex about the ridge line (with the lower boundary marking the transition to passive systems; Region 3). Horizontal dashed lines indicate the 3$\sigma$ and 5$\sigma$ radio detection limits converted to SFRs using the adopted radio–SFR calibration (equation~\ref{eq:cook}). The vertical line marks $M_\star =10^{10}\rm M_\odot$, above which the completeness criterion of equation~\ref{eq:thresh} is evaluated. The comparison illustrates the redshift-dependent ability of current and future radio surveys to unambiguously distinguish star-forming from passive SN Ia host galaxies.}
  \label{fig:mocksfms}
\end{figure}

In addition to wide-area coverage, dedicated SKAO continuum deep fields are anticipated, likely overlapping the Rubin/LSST DDFs and reaching an rms sensitivity of $S_{1.4} \sim 0.2\mu$Jy, roughly a factor of 20 deeper than our current deepest data on these fields \citep[see e.g.][]{2015aska.confE..67P,2015aska.confE..68J}. At $z=1$, the anticipated SKAO deep-field sensitivity places both the $3\sigma$ and $5\sigma$ detection limits below the SF-MS envelope for galaxies with $M_{\star} >10^{10}\,\rm M_\odot$ (bottom panel of Fig.~\ref{fig:mocksfms}), such that a radio non-detection would unambiguously correspond to a passive (Region~3) system. This continues to hold up to $z>2$, although, given the evolution in the SF-MS, there are a relative scarcity of passive galaxies at these redshifts.

It is worth noting that this depth (0.2 $\mu$Jy) far exceeds the equivalent depth for SFR of existing far-infrared data over these fields from \textit{Herschel} \citep{HerMES}, SCUBA2 \citep{Geach2017}, and WISE \citep{2010AJ....140.1868W}, making the radio continuum one of the most efficient probes of the total star-formation of a galaxy \citep[see e.g.][]{Smith2021,Cochrane2023,Thykkathu2026,2026arXiv260205808V}.

The principle advantage of SKAO over MeerKAT (whose MIGHTEE survey currently covers the DDFs), however, is not only its nominal sensitivity, but the fact that the SKAO will have sufficient angular resolution to not be predominantly impacted by the confusion limit. In contrast, the effective depth of MIGHTEE is limited by confusion noise at its $\sim5$~arcsec resolution, rather than by thermal noise, typically reducing the achievable sensitivity by a factor of $\sim2-3$. The future use of radio continuum surveys for separating SN host galaxies into star-forming and passive systems (Regions~2 and 3) will therefore be impactful over the next decade. This will be critical for fully exploiting the optical and near-infrared datasets from LSST and {\it Euclid} for SN cosmology.





\section{Conclusions} \label{sec:conclusions}

In this paper, we model the distribution of DES SN~Ia host galaxies on the SFR--$M_\star$ plane using SFRs calibrated from 1.4~GHz luminosities, and compare these results with our previous work based on SED-derived SFR measurements incorporating UV-to-FIR photometry. We identify three regions within the 1.4~GHz SFR--$M_\star$ plane, corresponding to low-mass hosts, high-mass star-forming hosts and high-mass passive hosts. Our main findings are as follows: 

\begin{enumerate}[label=(\roman*), leftmargin=*,align=left,labelindent=\parindent, labelsep=0.5em, itemindent=!]

\item Across all three regions, 84~per cent of SN~Ia host galaxies retain the same regional classification as in our analysis based on using full UV-to-FIR SED-derived SFRs.

\item   
The low-mass region shows the highest level of consistency between both analyses, with a 96~per cent classification agreement. This is expected, as stellar masses are primarily constrained by optical/NIR photometry in the SED-fitting, and these data remain unchanged between this work and our previous paper.    
\item Differences are primarily confined to the high-mass regime, reflecting the sensitivity of the star-forming--passive division on the adopted star-formation estimator. 
Nevertheless, the high-mass star-forming and passive populations show strong overall consistency between the radio- and SED-derived classifications, with  77~per cent and 87~per cent classification agreement, respectively.  
\item We also measure SN~Ia nuisance parameters ($\alpha, \beta, M$) for SNe in each host galaxy region of the 1.4 GHz SFR--$M_\star$ plane. The slope of the SN colour-luminosity relation ($\beta$) continues to show a dependence on the region of the SFR--$M_\star$ plane occupied by the SN host galaxies; SNe~Ia in high-mass, passive systems yield the lowest value of $\beta=2.41\pm0.22$ (a ${\sim}3.9\sigma$ difference relative to low-mass hosts). SNe~Ia in high-mass hosts are also brighter post-standardisation than those in low-mass hosts, with the largest magnitude difference of 0.095 mag observed between low-mass and high-mass passive hosts at the $3\sigma$ level. The statistical significances of some inter-region comparisons differ between the radio-based and SED-based host classifications, as expected given the redistribution of SNe between regions when adopting a different SFR tracer. Nevertheless, the inferred nuisance parameters from the two analyses agree within $\lesssim 1.1\sigma$ across all regions, consistent with statistical fluctuations. The persistence of region-dependent trends within the radio-defined host populations therefore provides independent confirmation for the environmental dependence of SN~Ia properties identified in our previous study.
\end{enumerate}

Historically, the limiting systematic for SN~Ia analyses has been photometric calibration. We are now entering a transitional phase in SN cosmology in which host-galaxy-dependent effects are emerging as the dominant contribution to the systematic uncertainty budget \citep{vincenzi2024darkenergysurveysupernova}. To model environmental dependencies, cosmological analyses typically implement a host galaxy stellar-mass-dependent correction. 
Stellar mass is favoured as the environmental proxy because it exhibits a measurable correlation with SN luminosities and for the practical advantage that it can be robustly estimated even when host photometric data are limited.  Despite its utility and widespread adoption, there is mounting evidence that this single-parameter treatment provides an incomplete description of the host-environment dependence of SN~Ia luminosities.  In particular, empirical trends between SN~Ia standardisation parameters and galaxy SFR suggest that additional environmental information is encoded beyond that captured by mass alone. This points toward the possibility that future iterations of the standardisation framework may require inclusion of SFR alongside galaxy mass.  

The implementation of this approach is fundamentally challenged by the need for accurate and internally consistent SFR measurements. Full UV-to-FIR SED fitting is widely regarded as one of the most reliable diagnostics of SFR. The required longer-wavelength coverage will not, however, be uniformly available across the entire footprints of forthcoming surveys such as LSST and Roman. This will preclude a fully self-consistent derivation of galaxy SFRs from SED-fitting alone, rendering the measurements dependent on spatial variations in wavelength coverage. 
Instead, robust characterisation of the star-formation activity of SN~Ia hosts will require a combined multi-tracer approach.

The radio continuum  provides one such option, offering a powerful and dust-unbiased tracer of star-formation activity. Of course, as with any tracer, there are several factors in its application that still require careful consideration. One being the choice of literature radio--SFR calibration (see Appendix~\ref{app:alt_radsfr}). Given that these region definitions are used to construct SN~Ia subsamples, any calibration-dependent mixing among them propagates directly into the inferred nuisance parameters. This dependence can, however, be incorporated into the cosmological analysis by treating the calibration choice as a methodological systematic. For example, in the absence of a uniquely preferred calibration, the cosmological fit can be evaluated under a set of plausible calibrations, and the resulting variation in recovered parameters folded into the overall uncertainty budget. Another consideration is the characteristic timescale probed by radio emission (of order ${\sim}100$ Myr), which differs from that of more instantaneous SFR (e.g., SED-based) tracers, and whose implications remain to be fully understood. Nevertheless, with charting the cosmic star-formation history as a central objective, the SKA will establish an era in which radio-derived SFRs--refined through sustained calibration efforts--become the de~facto standard in modern galaxy evolution studies \citep{2009ApJ...706..482M, 2015aska.confE..68J}. This work establishes a roadmap for integrating 1.4~GHz derived SFRs into SN~Ia host studies, enabling environmental characterisation commensurate with future SN experiments, and thereby unlocking their full cosmological potential.

Beyond the practical demands inherent to host galaxy classification at scale, the LSST era will likely also necessitate a reassessment of the current implementation of SN~Ia cosmology. One such shift will be the construction of environmentally defined sub-samples, for example, by restricting analyses to SNe~Ia in SF, low-mass hosts (as suggested by \citealp{2014MNRAS.445.1898C,2023MNRAS.519.3046K}) or to passive galaxies (as suggested by \citealp{2022ApJ...938...62C}). This shift will foreground new, previously secondary considerations, including uncertainties in host-galaxy classification arising from differences between choice of SFR estimator and/or calibration, contamination between environmental categories (e.g., passive hosts misidentified as star-forming), and the accurate modelling of novel selection effects. Considerations that, taken together, will reshape the structure of the systematic error budget, as even modest levels of environment misclassification  propagate into non-negligible biases in SN~Ia standardisation and cosmological inference. Against this backdrop, a central question remains unanswered: do our cosmological results vary significantly when derived from sub-samples of SNe~Ia occupying distinct host environments?


\section*{Acknowledgements}

The authors gratefully acknowledge Catherine Hale for their assistance. The authors additionally wish to thank the anonymous reviewer for their detailed comments, which helped to substantially improve the manuscript.
SR, MJJ, MV and IHW acknowledge support from the Oxford Hintze Centre for Astrophysical Surveys which is funded through generous support from the Hintze Family Charitable Foundation. MJJ also acknowledges the support of a UKRI Frontiers Research Grant [EP/X026639/1], which was selected by the European Research Council. MV is supported by UKRI Future Leaders Fellowship [UKRI2363]. MS acknowledges support from the Science and Technology Facilities Council (STFC) grant ST/Y001850/1.

\section*{Data Availability}

The DES-SN5YR data are all in the public domain \citep{2024ApJ...975....5S}\footref{desdr}. 
The MIGHTEE continuum DR1 data used in this work are released with \citet{2025MNRAS.536.2187H}.

\bibliographystyle{mnras}
\bibliography{example}


\appendix






\section{Confusion matrices with alternative radio--SFR calibrations} \label{app:alt_radsfr}

\begin{figure*} 
\centering
\begin{tabular}{cc}
\begin{subfigure}{0.48\textwidth}
\includegraphics[width=0.9\linewidth]{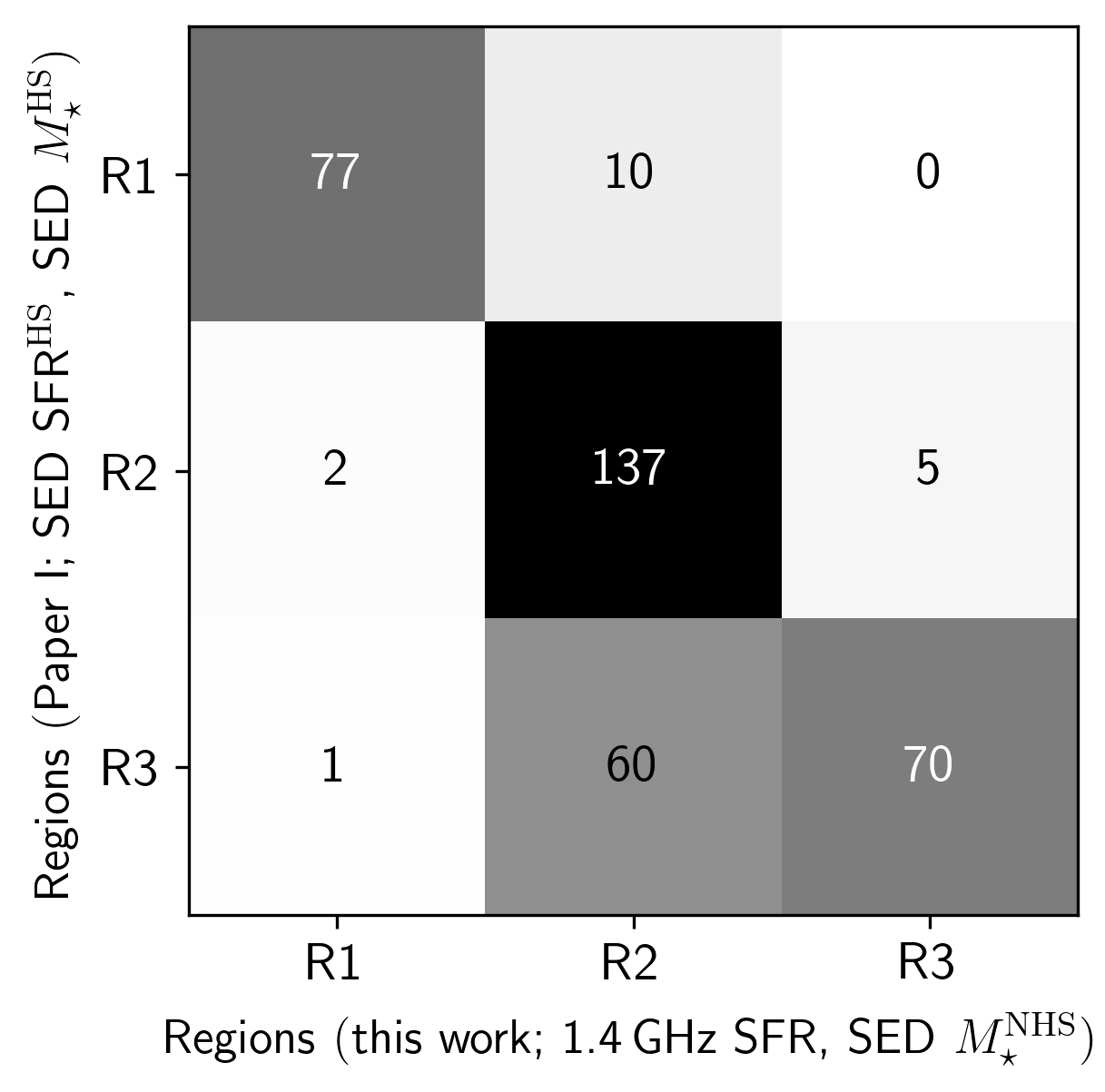}
\caption{\cite{1992ARA&A..30..575C} calibration.}
\end{subfigure} &
\begin{subfigure}{0.48\textwidth}
\includegraphics[width=0.9\linewidth]{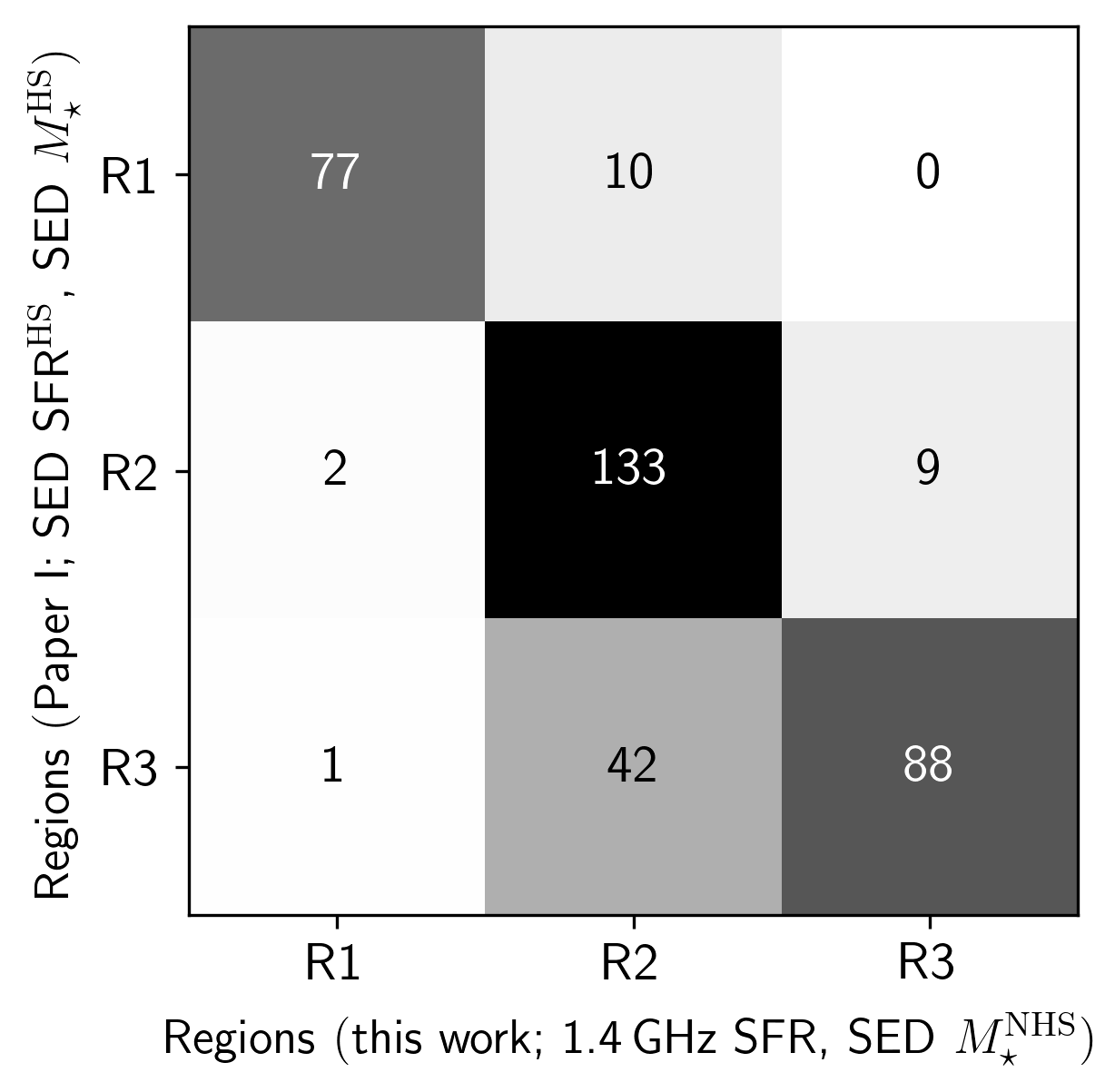}
\caption{\cite{2003ApJ...586..794B} calibration.}
\end{subfigure} \\
\begin{subfigure}{0.48\textwidth}
\includegraphics[width=0.9\linewidth]{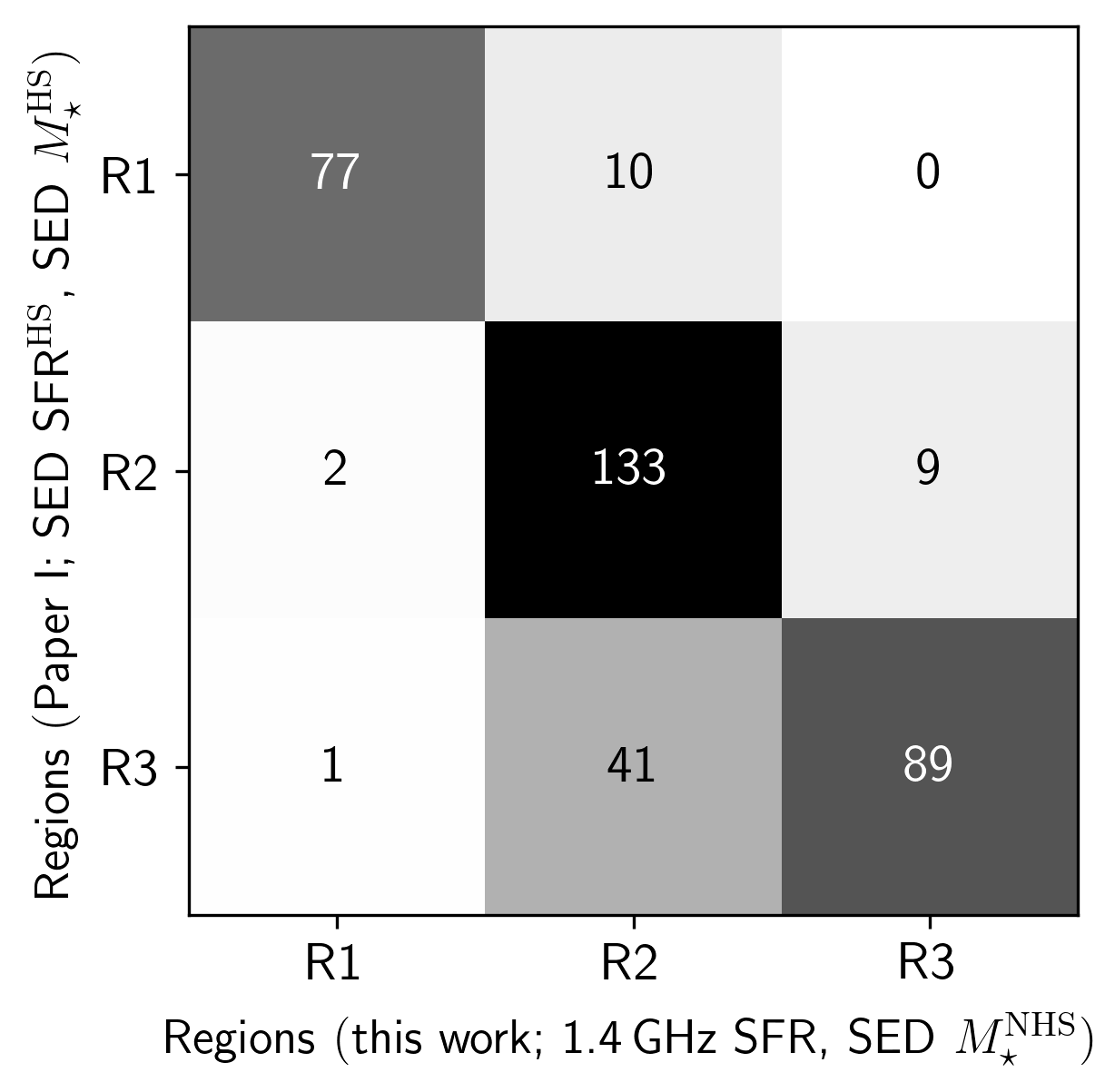}
\caption{\cite{2017MNRAS.466.2312D} calibration.}
\end{subfigure} &
\end{tabular}
\caption{Confusion matrices comparing the galaxy-region classifications from Paper~I (rows), based on SED-derived properties including \textit{Herschel} and \textit{Spitzer} data,  with those obtained in this work (columns) using 1.4~GHz SFRs. As in Fig.~\ref{fig:confusionm}, but with radio-derived SFRs computed using the \protect\cite{1992ARA&A..30..575C,2003ApJ...586..794B} and \protect\cite{2017MNRAS.466.2312D} calibrations in place of the \protect\cite{2024MNRAS.531..708C} relation used in our fiducial analysis. Cell values give the number of galaxies in each pairing.}
\label{fig:othercm}
\end{figure*}

As a robustness check, we explore the impact of radio--SFR calibration choice on galaxy-region classifications by comparing   confusion matrices for the \cite{1992ARA&A..30..575C, 2003ApJ...586..794B} and \cite{2017MNRAS.466.2312D} prescriptions, shown in Fig.~\ref{fig:othercm}. The overall structure of the confusion matrices is preserved, although there is a systematic increase in off-diagonal scatter relative to the fiducial case \citep{2024MNRAS.531..708C}, most notably for the \cite{1992ARA&A..30..575C} relation. These differences are consistent with the distinct assumptions underlying each relation. In particular, the calibrations are normalised using different definitions of SFR and treatments of the radio emission components, while the \cite{2017MNRAS.466.2312D} relation is empirically tied to SED-based SFRs (as used in this work) but derived from a relatively small sample. As such, variations at this level are expected when applying these prescriptions to  different datasets. We note that the \cite{2024MNRAS.531..708C} calibration adopted in the main analysis (Fig.~\ref{fig:confusionm}) was chosen prior to performing these comparisons and  not in an attempt to maximise agreement between classification schemes.



\bsp	
\label{lastpage}
\end{document}